\PassOptionsToPackage{numbers,sort&compress}{natbib}
\documentclass[preprintnumbers,amsmath,amssymb,prd, notitlepage,nofootinbib,twocolumn, superscriptaddress]{revtex4}

\usepackage{amsfonts,amssymb,amsmath}
\usepackage{gensymb}
\usepackage{color}
\usepackage{graphicx}
\usepackage{multirow}
\usepackage[utf8]{inputenc}
\usepackage{aas_macros}
\usepackage{hyperref}
\usepackage{xcolor}
\usepackage[a4paper, total={6.5in, 9in}]{geometry}

\graphicspath{{./Images/}}
\usepackage[normalem]{ulem}

\newcommand{\greaterthanapprox}{\mathrel{\vcenter{
  \offinterlineskip\halign{\hfil$##$\cr
    >\cr\noalign{\kern2pt}\sim\cr\noalign{\kern-2pt}}}}}
    
    \newcommand{\lessthanapprox}{\mathrel{\vcenter{
  \offinterlineskip\halign{\hfil$##$\cr
    <\cr\noalign{\kern2pt}\sim\cr\noalign{\kern-2pt}}}}}

\newcommand{\Planck}{{\it Planck}~}

\newcommand{\be}{\begin{equation}}        
\newcommand{\ee}{\end{equation}}

\newcommand{\fnl}[0]{$f_\mathrm{NL}$}

\newcommand\numberthis{\addtocounter{equation}{1}\tag{\theequation}}

\usepackage[normalem]{ulem}
\usepackage{xcolor}

\newcommand{\pypackage}[1]{\texttt{#1}}
\newcommand{\software}[1]{\textsc{#1}}

\defcitealias{mccarthyConstraintsPrimordialNonGaussianity2023}{McC23}
\newcommand{\McCfNL}{\citetalias{mccarthyConstraintsPrimordialNonGaussianity2023}}
\defcitealias{mccarthyLargescaleGalacticdustcleanedCosmic2024}{McC24}
\newcommand{\McCCIB}{\citetalias{mccarthyLargescaleGalacticdustcleanedCosmic2024}}
\defcitealias{lenzLargescaleMapsCosmic2019}{L19}
\newcommand{\Lenz}{\citetalias{lenzLargescaleMapsCosmic2019}}
\begin{document}

\title{New constraints on primordial non-Gaussianity from large-scale cross-correlations of CMB lensing and the cosmic infrared background}

\author{Joseph Thornton}
\email{jdt50@cam.ac.uk}
\affiliation{DAMTP, Centre for Mathematical Sciences, Wilberforce Road, Cambridge CB3 0WA, UK}

\author{Fiona McCarthy}
\affiliation{DAMTP, Centre for Mathematical Sciences, Wilberforce Road, Cambridge CB3 0WA, UK}
\affiliation{Kavli Institute for Cosmology Cambridge, Madingley Road, Cambridge, CB3 0HA, UK}
\affiliation{Center for Computational Astrophysics, Flatiron Institute, New York, NY, USA 10010}

\author{Carmen Embil Villagra}
\affiliation{DAMTP, Centre for Mathematical Sciences, Wilberforce Road, Cambridge CB3 0WA, UK}
\affiliation{Kavli Institute for Cosmology Cambridge, Madingley Road, Cambridge, CB3 0HA, UK}

\author{Blake D. Sherwin}
\affiliation{DAMTP, Centre for Mathematical Sciences, Wilberforce Road, Cambridge CB3 0WA, UK}
\affiliation{Kavli Institute for Cosmology Cambridge, Madingley Road, Cambridge, CB3 0HA, UK}

\date{\today}

\begin{abstract}

We present new constraints on the local-type primordial non-Gaussianity parameter, $f_\mathrm{NL}^\mathrm{local}$, through analysis of the scale-dependent bias effect on the cosmic infrared background (CIB).
To avoid biases from galactic dust contamination on large scales, we use cross-correlations between the CIB and \Planck cosmic microwave background (CMB) lensing maps to constrain non-Gaussianity.
Our measurement employs new dust-cleaned CIB maps that have been designed to be unbiased on large scales, which allows us to improve our constraining power on $f_\mathrm{NL}^\mathrm{local}$ by a factor of $\sim 2$ over previous CIB analyses.
We derive a constraint of $f_\mathrm{NL}^\mathrm{local}=43 \pm 23$, matching the precision of the tightest existing constraints from cross-correlation methods.
Consistency- and null-tests demonstrate that our results are robust to modeling assumptions and residual dust contamination.

\end{abstract}
\maketitle

\section{Introduction}
Inflation \cite{GuthInflationaryuniverse1981} is a proposed initial accelerated phase of expansion which not only addresses several problems inherent in the standard hot big bang cosmology, but also provides a mechanism to explain the origin of structure in the universe.
Despite its success, the specific nature or number of its sourcing fields is unknown.
Observables that can provide information about this field are therefore of crucial importance in constraining fundamental physics.
One aspect that allows the distinguishing of different models is the imprint of non-Gaussianity that they leave on observables.
It has been argued~\cite{deputterNextNonGaussianityFrontier2017} that multi-field inflation generically gives rise to local-type primordial non-Gaussianity (PNG).
This is expressed in the typical form~\cite{komatsuAcousticSignaturesPrimary2001}:
\begin{equation}
    \phi(\mathbf x) = \phi_G(\mathbf x) + f^\mathrm{local}_{\mathrm{NL}}\left(\phi_G(\mathbf x)^2 - \left\langle\phi_G^2\right\rangle\right) \label{eq:fNL},
\end{equation}
where $\phi$ is the primordial potential,  $\phi_G$ is a single Gaussian field, and $\mathbf x$ refers to a vector of spatial coordinates.
As such, $f_\mathrm{NL}^\mathrm{local}$ parametrizes the level of local-type non-Gaussianity of the potential.
A detection of $f_\mathrm{NL}^\mathrm{local}\sim\mathcal{O}(1)$ would rule out single-field slow-roll inflation, which predicts a neglible level of~$f_\mathrm{NL}^\mathrm{local}$~\cite{2013PhRvD..88h3502P,2015JCAP...10..024D}.
Conversely, inflation driven by several interacting scalar fields predicts $f_\mathrm{NL}^\mathrm{local}\sim\mathcal{O}(1)$.
The current best constraint on the local-type PNG parameter come from the \textit{Planck} Collaboration analysis of the cosmic microwave background (CMB) bispectrum~\cite{planckcollaborationPlanck2018Results2020}, with $f_\mathrm{NL}^\mathrm{local}=-0.9 \pm 5.1$ from the official \textit{Planck} PR3 analysis and $f_\mathrm{NL}^\mathrm{local}=-0.1 \pm 5.0$  from an updated PR4 analysis~\cite{2025A&A...702A.204J}.\footnote{Excitingly, large scale structure constraints have very recently reached a  point where they are comparable enough with this measurement to be meaningfully combined with it, such that Ref.~\cite{2025arXiv251204266C}  finds $f_\mathrm{NL}^\mathrm{local}=-0.0 \pm 4.1$  from the combination with DESI DR1 galaxies~\cite{2026AJ....171..285D}, and Ref.~\cite{2026arXiv260405213R} would find a similar errorbar upon such a combination. } 

In addition to the bispectrum, the local-type PNG is also observable in the power spectrum of biased tracers of matter density.
The presence of non-zero local PNG induces a specific scale-dependence  to the bias~\cite{dalalImprintsPrimordialNongaussianities2008} such that
\begin{equation}
    \Delta b \propto \frac{f_\mathrm{NL}}{k^2},
\end{equation}
where $k$ indicates comoving wavenumber, $\Delta b$ is the additional bias due to PNG, and where we now drop the `local' superscript for the duration of the paper.

A number of studies have constrained \fnl{} through the scale-dependent bias of different tracers (e.g.~\cite{SlosarConstraintsonlocal2008,2013MNRAS.428.1116R,2014PhRvL.113v1301L,2014PhRvD..89b3511G,
2014MNRAS.441L..16G,2022MNRAS.514.3396M,mccarthyConstraintsPrimordialNonGaussianity2023,2024JCAP...03..021K,2024MNRAS.532.1902R,chaussidonConstrainingPrimordialNonGaussianity2024, FabbianConstraintsonprimordial2025, 2024arXiv240805264K,2025PhRvL.134o1003L,HotinliVelocityReconstructionfrom202, 2025arXiv251115701M,chiarenza_constraining_2025,2026arXiv260116948R,2026arXiv260404867C,2026arXiv260405213R}) or the combination of this signal with the galaxy bispectrum~\cite{2022PhRvD.106d3506C,2025arXiv251204266C}, with the current tightest constraints coming from DESI DR1~\cite{chaussidonConstrainingPrimordialNonGaussianity2024,2025arXiv251204266C,2026arXiv260405213R}, with~$1\sigma$ credible intervals, $\sigma(f_\mathrm{NL})\approx 9$.
In this work we constrain \fnl{} through analysis of scale-dependent bias between the cosmic infrared background (CIB) and matter, which we constrain by measuring the cross-correlation of the CIB and the CMB lensing potential.

The CMB~\cite{1965ApJ...142..419P,1965ApJ...142..414D} is the oldest light in the universe, sourced at the surface of last scattering, at redshift $z\sim 1100$.
As this light travels towards us through cosmic time, gravitational interactions with intervening matter cause weak deflections from its original trajectory~\cite{1987A&A...184....1B}. 
This weak `lensing' distorts the CMB, and the potential that sources this can be estimated from measurements of the temperature and polarization anisotropies.
The lensing potential is sourced by the projected matter density along a given line of sight, and a reconstruction of the lensing potential is thus an unbiased tracer of the projected matter field.
The kernel for CMB lensing is wide in redshift space, but has its peak around $z\sim 2$.
See~\cite{2006PhR...429....1L} for a review of CMB lensing.
The CMB lensing maps used for this work are the \Planck{} NPIPE CMB lensing maps of Ref.~\cite{carronCMBLensingPlanck2022a}.

The CIB~\cite{1996A&A...308L...5P} is a full-sky field primarily sourced by dust in star-forming galaxies that has been heated by the ultra-violet radiation of young stars.
It is sourced across cosmic time, but peaks around the ``cosmic noon'' at redshifts $z\sim2$~ \cite{MadauDickinsonCosmicStar2014}.
It can act as both an integrated probe of star-formation history, and as a biased tracer of large-scale structure, the latter point being key to its use here in PNG constraints.
It is very sensitive to \fnl{} due to its high-redshift origin, since bias generally increases with redshift.
In principle, constraints on~\fnl{} could be made in both auto-correlations \cite{tucciCosmicInfraredBackground2016} and cross-correlations.
However, CIB maps are contaminated by Galactic dust that is hard to clean on large-scales.
The auto-correlation of this foreground contamination adds power to the low-$\ell$ end of the CIB auto-spectra, which can mimic the effect of \fnl.
If one could perfectly clean foreground dust from the CIB, use of the auto-spectrum for similar data and analysis choices to this work are forecast to provide constraints on $\sigma(f_\mathrm{NL})\sim 2$~\cite{mccarthyConstraintsPrimordialNonGaussianity2023}.
In principle, future surveys with access to larger angular scales and with perfect dust cleaning could reach~$\sigma(f_\mathrm{NL})<1$~\cite{tucciCosmicInfraredBackground2016}.

Generally, the cross-correlations of biased tracers are a powerful and robust avenue for constraining~\fnl, as they can help mitigate systematics and contaminants that affect individual tracers.
The CIB is highly correlated with the CMB lensing potential~\cite{2003ApJ...590..664S,Planck2013XVIII}, as both are sourced by the same underlying matter distribution and have broad redshift kernels peaking around similar redshifts.
Since CMB lensing is minimally affected by Galactic dust~\cite{abril-cabezasImpactGalacticNonGaussian2025} and the CIB is highly sensitive to \fnl{}, their cross-correlation is particularly well-suited for \fnl{} studies.
Because most of the \fnl{} signal resides on large scales, we restrict our large-scale analysis to CIB-CMB lensing cross-correlations, using CIB auto- and inter-frequency spectra only on intermediate scales to constrain bias and redshift evolution.

Constraints have been placed before on \fnl{} through CIB-CMB lensing cross-correlations in Ref.~\cite{mccarthyConstraintsPrimordialNonGaussianity2023} (hereafter referred to as~\McCfNL), which used CIB maps produced in Ref.~\cite{lenzLargescaleMapsCosmic2019} (hereafter referred to as \Lenz).
The cleaning procedure used meant these maps were biased below multipoles of~$\ell\approx 70$ (i.e., the cosmological signal was oversubtracted on these scales), and as the scale-dependence is strongest at large scales, we wish to push lower for maximum constraining power.
In this work we use the maps produced in Ref.~\cite{mccarthyLargescaleGalacticdustcleanedCosmic2024} (hereafter referred to as~\McCCIB), {which were cleaned with the goal of being unbiased on larger scales than those of \Lenz{} (See Section~\ref{sec:data_cib} for more details).}
These new maps are unbiased down to much lower $\ell$; in this work we use a minimum multipole $\ell_\mathrm{min}=19$.

Galaxy clustering analyses face similar problems to the CIB: large-scale systematics can make the signal hard to measure in galaxy clustering, and so cross-correlation-alone analyses have been used to constrain \fnl{} in correlation with CMB lensing ~\cite{2024JCAP...03..021K,FabbianConstraintsonprimordial2025,chiarenza_constraining_2025} as well as with other unbiased tracers such as the integrated Sachs--Wolfe (ISW) effect~\cite{2014MNRAS.441L..16G,2014PhRvD..89b3511G,2026arXiv260116948R}, and most recently the large-scale velocity field reconstructed with kinematic Sunyaev--Zel'dovich (kSZ) measurements~\cite{1980MNRAS.190..413S,2018PhRvD..98l3501D,2018arXiv181013423S,2019PhRvD.100h3508M,2024arXiv240805264K,2025PhRvL.134o1003L,HotinliVelocityReconstructionfrom202,2025arXiv251115701M,2026arXiv260404867C}.
The current tightest constraints from cross-correlations alone have errorbars~$\sigma(f_{\mathrm{NL}})\sim22$~\cite{chiarenza_constraining_2025,2026arXiv260116948R}.
More generally, including a cross-correlation in an analysis can also improve upon auto-correlation-only constraints due to sample variance cancellation~\cite{2009PhRvL.102b1302S,2018PhRvD..97l3540S}.

The remainder of this paper is organized as follows. In Section~\ref{sec:theory} we discuss the theoretical models used for this analysis.
In Section~\ref{sec:data} we discuss the data products used in more detail.
In Section~\ref{sec:likelihood} we discuss our likelihood, as well as covariance estimation and simulations.
We present our results in Section~\ref{sec:results}, and our conclusions in Section~\ref{sec:conclusions}.

For all cosmological models we use \Planck 2018 cosmological parameters~\cite{PlanckCosmologicalParameters2020} \{$H_0=67.32\, \mathrm{km\, s^{-1}\,Mpc^{-1}}$, $A_s = 2.1\times 10^{-9}$, $n_s = 0.9661$, $\Omega_ch^2 = 0.12011$, $\Omega_bh^2 = 0.022383$\}, where $H_0$ is the value of the Hubble parameter today; $A_s$ is the amplitude of scalar perturbations at a pivot scale of $0.05\, \mathrm{Mpc}^{-1}$; $n_s$ is the scalar spectral index; and $\Omega_c h^2$ and $\Omega_b h^2$ are the physical densities of cold dark matter and baryons respectively.
\section{Theory}\label{sec:theory}

In this Section, we provide the background theory required to model our signal.
In Section~\ref{sec:scaledepbias} we present the scale-dependent effect of \fnl on linear bias.
In Section~\ref{sec:angularpowerspectra} we present the calculation of the theoretical angular cross power spectrum between two projected fields, which we specify as CMB lensing in Section~\ref{sec:cmblensingtheory} and the CIB in Section~\ref{sec:cib_model}.
In Section~\ref{sec:instrumentaleffectsmodeling} we present some details of instrumental effects that we model.
\subsection{Scale-dependent bias}\label{sec:scaledepbias}

Non-Gaussianity is often parameterized through~\fnl, the coefficient in the second-order perturbative expansion of the primordial potential,~$\phi$, as a function of a single Gaussian field,~$\phi_G$, as shown in Equation~\eqref{eq:fNL}.
Non-Gaussianity of this form introduces an additional $k^{-2}$ scale dependent term to the linear bias of biased tracers of the matter field~\cite{dalalImprintsPrimordialNongaussianities2008}.
Here, the linear bias~$b$ is defined as the proportionality between the overdensity~$\delta_\mathrm{tracer}$, of a tracer, and the matter overdensity~$\delta_m$, i.e.~$\delta_{\mathrm{tracer}} = b \delta_m$ on large scales ($b$ is tracer-dependent).
Explicitly,
\begin{equation}
    b^{NG} = b^G + f_\mathrm{NL}\frac{3\Omega_m H_0^2}{2k^2 T(k) D(z)}b_{\phi} ,\label{eq:bNG}
\end{equation}
where $b^G$ is the Gaussian bias and $b^{NG}$ is the full non-Gaussian bias; $\Omega_m$ is the matter density today in units of the critical density; $H_0$ is the Hubble constant; $T(k)$ is the matter transfer function,  normalized to $1$ at low $k$; $D(z)$ is the growth function, normalized to $\frac{1}{1+z}$ during matter domination; and $b_{\phi}$ is the local PNG bias which describes the response of the biased field to local PNG. There is an important degeneracy apparent then between $b_\phi$ and \fnl{} that will need to be settled in order to truly interpret \fnl{} constraints at more than an $\mathcal{O}(1)$ level~\cite{barreiraCanWeActually2022a}.
In this work, the universality relation is assumed between~$b^G$ and $b_\phi$ of $b_\phi = 2\delta_c(b^G - 1)$ \cite{SlosarConstraintsonlocal2008}, where $\delta_c=1.686$ is the critical density for spherical collapse, giving a final change to the Gaussian bias as
\begin{equation}
    \Delta b = f_\mathrm{NL}\frac{3\Omega_mH_0^2}{k^2T(k)D(z)}\delta_c(b^G - 1)\label{eq:delta_bNG}.
\end{equation}

\subsection{Angular cross-power spectra}\label{sec:angularpowerspectra}

Consider a projected field $F^{2D}(\hat{\mathbf{n}})$, where $\mathbf{\hat{n}}$ is the line of sight direction..
This projected field is produced by the projection of some  3D field,~$\alpha^{3D}(\chi, \hat{\mathbf{n}})$, where $\chi$ is the comoving distance.
The projected field can then be written as
\begin{equation}
    F^{2D}(\mathbf{\hat{n}}) = \int \mathrm{d}\chi W^F(\chi)\alpha^{3D}(\chi,\mathbf{\hat{n}})\label{eq:projected_field},
\end{equation}
where $W^F(\chi)$ is the redshift kernel of the projected field, which may encode intrinsic properties of the projected field $F^{2D}$, or properties of a given survey.
This projected field can be written as spherical harmonic expansion with
\begin{equation}
    F^{2D}(\mathbf{\hat{n}}) = \sum_{\ell=0}^\infty \sum_{m=-\ell}^\ell F_{\ell m}Y_\ell^m(\mathbf{\hat{n}})\label{eq:spherical_harmonic_expansion},
\end{equation}
where $F_{\ell m}$ are the spherical multipole coefficients, and $Y_\ell^m(\mathbf{\hat{n}})$ are the spherical harmonic functions.
Through Fourier transform of $\alpha^{3D}(\chi,\mathbf{\hat{n}})$ and plane-wave expansion, one can express these coefficients as
\begin{align*}
    F_{\ell m} = \frac{1}{2\pi^2}&\int \mathrm{d}\chi W^F(\chi)\\
    &\int\mathrm{d}\mathbf{k}\alpha^{3D}(\mathbf{k}, z(\chi))i^\ell j_\ell(k\chi)Y_\ell^{m*}(\mathbf{\hat{k}})\label{eq:spherical_harmonic_coefficients},
    \numberthis
\end{align*}
where $\alpha^{3D}(\mathbf{k},z(\chi))$ is the Fourier transform of the field  $\alpha^{3D}(\chi,\mathbf{\hat{n}})=\int\frac{\mathrm{d}\mathbf{k}}{(2\pi)^3}\alpha^{3D}(\mathbf{k}, z(\chi))e^{i\mathbf{k}\cdot\chi\mathbf{\hat{n}}}$ (here $z$ plays the role of our time coordinate);   $j_\ell(k\chi)$ are the spherical Bessel functions of degree $\ell$; the star indicates the complex conjugate (of the spherical harmonics);  and the argument of $Y_\ell^{m*}$ is the direction of $\mathbf{k}$ such that $\mathbf{k}\equiv k \mathbf{\hat{k}}$.
For two such fields~$F^{2D}$ and~$G^{2D}$ (which are statistically isotropic), produced from projections of 3D fields $\alpha^{3D}$ and $\beta^{3D}$ respectively, the angular power spectrum $C_\ell^{FG}$ is defined as
\begin{equation}
	\left\langle F_{\ell m} G_{\ell' m'}^*\right\rangle \equiv C_\ell^{FG}\delta_{\ell\ell'}\delta_{mm'}\label{eq:angular_power_def},
\end{equation}
where the expectation $\langle\cdot\rangle$ is over realizations of the fields.
From Equation~\eqref{eq:spherical_harmonic_coefficients} we can therefore show that
\begin{align*}
    C_\ell^{FG} = &\frac{2}{\pi}\int \mathrm{d}\chi \mathrm{d}\chi'\int k^2 \mathrm{d}k \\\numberthis
    &W^F(\chi) W^G(\chi')P_{\alpha \beta}(k, z, z')j_\ell(k\chi)j_\ell(k\chi'),\label{eq:general_angular_full}
\end{align*}
where $P_{\alpha\beta}$ is the power spectrum of $\alpha^{3D}$ and $\beta^{3D}$ defined as $\langle\alpha^{3D}(\mathbf{k}, z)\beta^{*3D}(\mathbf{k'}, z')\rangle \equiv P_{\alpha\beta}(k, z, z')\delta(\mathbf{k}-\mathbf{k'})$.

If the fields in question are biased tracers of the matter field, $m$, and we assume a linear bias model, we have 
\begin{align}
    P_{\alpha m}(k, z, z') &= b_\alpha(z) P^\mathrm{lin}_{mm}(k, z, z');\\
    P_{\alpha\beta}(k, z, z') &= b_\alpha(z) b_\beta(z') P^\mathrm{lin}_{mm}(k, z, z'),
\end{align}
where $P^\mathrm{lin}_{mm}$ is the linear matter power spectrum,~$P_{\alpha m}$ is the cross-power spectrum of $\alpha^{3D}$ and matter, and $b_{\alpha/\beta}$ indicates the linear bias of the field~$\alpha^{3D}$/$\beta^{3D}$.
{From Equation~\eqref{eq:general_angular_full} we can then define the angular power spectrum of this linear bias model to be}
\begin{align*}
    C_{\ell\mathrm{,lin}}^{FG} = &\frac{2}{\pi}\int \mathrm{d}\chi \mathrm{d}\chi'\int k^2 \mathrm{d}k j_\ell(k\chi)j_\ell(k\chi')\\\numberthis
    &b_\alpha(\chi) b_\beta(\chi') W^F(\chi) W^G(\chi')P^\mathrm{lin}_{mm}(k, z, z').\label{eq:general_angular_linear_bias}
\end{align*}
For broad redshift kernels (except on very large scales) we may then also employ the Limber approximation~\cite{limberAnalysisCountsExtragalactic1953}, in which case Equation~\eqref{eq:general_angular_linear_bias} reduces to 
\begin{align*}
    C_{\ell\mathrm{,lin}}^{FG} = &\int \frac{\mathrm{d}\chi}{\chi^2}\\\numberthis
    &b_\alpha(\chi) b_\beta(\chi) W^F(\chi) W^G(\chi)P^\mathrm{lin}_{mm}\left(k = \frac{\ell}{\chi}, z\right).\label{eq:general_angular_limber}
\end{align*}
On all scales used in this work, the Limber approximation is adequate.
We also include a shot-noise and $1$-halo term at the angular power spectrum level of the CIB auto-correlations (see Section~\ref{sec:cib_model}).
These terms only begin to impact the theory spectra at the smaller scales of our analysis.

All calculations of the linear matter power spectrum and background cosmology are performed with \texttt{CAMB}\footnote{\url{https://github.com/cmbant/camb}}~\cite{2011ascl.soft02026L}.
\subsubsection{CMB Lensing}\label{sec:cmblensingtheory}
The primary CMB is sourced at $z\sim 1100$, at the surface of last scattering.
On its path to us, it interacts with intervening matter, giving rise to `secondary anisotropies'.
One such secondary is CMB lensing, where gravitational potentials along the line of sight weakly deflect CMB photons.
In a flat Universe, the lensing potential $\phi(\hat{\mathbf n})$ is given by (see, e.g.~\cite{2006PhR...429....1L})
\begin{equation}
    \phi(\hat{\mathbf n}) = -2\int_0^{\chi_*}\mathrm{d}\chi\frac{\chi_* - \chi}{\chi_*\chi}\Phi(\chi, \hat{\mathbf{n}}),\label{eq:lensing_potential}
\end{equation}
where $\chi_*$ is the comoving distance to the source of the lensed photons, which for  CMB lensing is the surface of last scattering.
Here, $\Phi(\chi, \hat{\mathbf{n}})$ denotes the Newtonian potential, which is related to the matter overdensity $\delta$ on sub-horizon scales by 
\begin{equation}
    \nabla^2\Phi(\chi, \hat{\mathbf{n}}) = \frac32\left(\frac{H_0}{c}\right)^2\frac{\Omega_m}{a(\chi)}\delta(\chi, \hat{\mathbf{n}}),\label{eq:newtonian_potential}
\end{equation}
where $c$ is the speed of light, and $a(\chi)$ is the scale factor at $\chi$.
The convergence field, $\kappa$, is related to the lensing potential through the relationship $\nabla^2\phi = -2\kappa$.
We can therefore express the convergence field as
\begin{equation}
    \kappa(\hat{\mathbf{n}}) = \int_0^{\chi_*}\mathrm{d}\chi\frac32\left(\frac{H_0}{c}\right)^2\frac{\Omega_m}{a(\chi)}\chi\left(1-\frac{\chi}{\chi_*}\right)\delta(\chi, \mathbf{\hat{n}}),
\end{equation}
from which we can see that the lensing convergence kernel is given by 
\begin{equation}
    W^\kappa(\chi) = \frac32\left(\frac{H_0}{c}\right)^2\frac{\Omega_m}{a(\chi)}\chi\left(1-\frac{\chi}{\chi_*}\right)\label{eq:converngence_kernel}.
\end{equation}
\subsubsection{CIB}\label{sec:cib_model}
The CIB intensity, $I_\nu$, at a given frequency, $\nu$, is given by
\begin{equation}
    I_\nu(\mathbf{\hat{n}}) = \int \mathrm d\chi \,a(\chi)j_\nu(\chi, \mathbf{\hat{n}}) \label{eq:CIB_intensity},
\end{equation}
where $j_\nu(\chi, \mathbf{\hat{n}})$ is the comoving emissivity density.
By breaking the emissivity density into $\overline{j}_\nu(\chi)$ and~$\delta j_\nu(\chi, \hat{\mathbf n})$, its mean and fluctuations respectively, we can consider the emissivity power spectrum, $P^{\nu\nu'}_{jj}$, as (see, e.g.~\cite{2001ApJ...550....7K,2012MNRAS.421.2832S})
\begin{equation}
    \frac{\langle \delta j_\nu(\mathbf k, z)\delta j_\nu(\mathbf k', z')\rangle}{\overline{j}_\nu(z)\overline{j}_{\nu'}(z')} \equiv (2\pi)^3 P^{\nu\nu'}_{jj}(k, z, z')\delta(\mathbf k -\mathbf k')\label{eq:emmisivity_power}.
\end{equation}
As mentioned in Section~\ref{sec:angularpowerspectra}, on the large angular scales of interest, we can neglect non-linear clustering and assume a linear bias model for the emissivity power spectrum.
As such we relate our emissivity power spectrum to the linear matter power spectrum through a linear bias $b_{\mathrm{CIB}}(z)$
\begin{equation}
    P_{jm}(k, z, z') = b_\mathrm{CIB}(z)P^\mathrm{lin}_{mm}(k, z, z')\label{eq:emmisivity_to_matter}.
\end{equation}
Relating all of this to Equations~\eqref{eq:projected_field} and \eqref{eq:general_angular_limber}, we see that the redshift kernel of the CIB intensity at a frequency $\nu$, $W^\nu(\chi)$, is given by
\begin{equation}
    W^\nu(\chi) = a(\chi) \overline{j}_\nu(\chi)\label{eq:nu_kernel},
\end{equation}
with the linear bias entering Equation~\eqref{eq:general_angular_limber} being $b_\mathrm{CIB}(z)$.
For modeling $b_\mathrm{CIB}$ and $\overline{j}_\nu(z)$ we  use the parametrization outlined in Ref.~\cite{maniyarHistoryStarFormation2018a} (this is the same model that was used in \McCfNL), where
\begin{equation}
    b_\mathrm{CIB}(z) = b_0 + b_1 z + b_2 z^2,\label{eq:bCIB}
\end{equation}
with $b_0, b_1, b_2$ being three free parameters.
In Section~\ref{sec:alternative_bias} we explore alternative parameterizations to test that our results are robust to changes to the bias model.

The mean emissivity is related to the mean star formation rate density (SFRD), $\rho_\mathrm{SFR}(z)$, by the Kennicutt relation \cite{kennicuttStarformationin1998}
\begin{equation}
    \overline{j}_\nu(z) = \frac{1}{K}\left[\rho_\mathrm{SFR}(z)(1+z)S_{\nu, \mathrm{eff}}(z)\chi(z)^2\right], \label{eq:kennicutt}
\end{equation}
where $K=1\times10^{-10}M_\odot\mathrm{yr}^{-1}L_\odot^{-1}$ is the Kennicutt constant and $S_{\nu, \mathrm{eff}}(z)$ is the effective spectral energy distribution (SED).
We use the same SEDs as those used in Ref.~\cite{maniyarHistoryStarFormation2018a}, which were calculated from \textit{Herschel} data \cite{betherminEvolutionDustEmission2015, betherminImpactClusteringAngular2017} with the method outlined in Ref.~\cite{betherminRedshiftEvolutionDistribution2013}.
The SFRD is parameterized as
\begin{equation}
    \rho_\mathrm{SFR}(z) = \frac{\alpha (1+z)^\beta}{1 + \left(\frac{1+z}{\gamma}\right)^\delta},
\end{equation}
giving an additional four free parameters, $\alpha, \beta, \gamma, \delta$, based on the parameterization suggested in Ref.~\cite{MadauDickinsonCosmicStar2014}.
To include \fnl{} we promote Equation~\eqref{eq:bCIB} to the non-Gaussian bias by adding $\Delta b$ from Equation~\eqref{eq:delta_bNG}, where $b_G=b_\mathrm{CIB}$.

As mentioned in Section~\ref{sec:angularpowerspectra}, a purely linear model is not sufficient to describe the higher-$\ell$ portion of our CIB auto-spectra. As such, we use a halo model prescription to add some non-linear power (which is subdominant).
The halo model (see~\cite{2002PhR...372....1C} for a review) is a framework that describes the non-linear distribution of matter by assuming all mass resides in gravitationally bound dark matter halos.
At the two-point level, it decomposes matter correlations into intra-halo correlations (1-halo term) and inter-halo correlations (2-halo term), providing a description of structure formation across scales.
While we are not using a halo-model throughout, we use it to account for a small contribution from the 1-halo term.
We also include a contribution from shot-noise, which arises as a result of the CIB being produced by discrete, albeit unresolved Poisson-distributed sources.

This results in our model for the final angular power spectra of the CIB {intensity}, $C_\ell^{\nu\nu'}$, being
\begin{equation}
    C_\ell^{\nu\nu'} = C_{\ell, \mathrm{lin}}^{\nu\nu'} + C_{\ell, \mathrm{1h}}^{\nu\nu'} + S_{\nu, \nu'},\label{eq:full_nunup}
\end{equation}
where $C_{\ell, \mathrm{lin}}^{\nu\nu'}$ arises from the projection of the linear term described in this section, $ C_{\ell, \mathrm{1h}}^{\nu\nu'} $ is the one-halo term, and $ S_{\nu, \nu'}$ is the shot noise term.
Following Ref.~\cite{maniyarHistoryStarFormation2018a} and \McCfNL{} we include precomputed fiducial values for $C_{\ell, \mathrm{1h}}^{\nu\nu'}$ and $S_{\nu, \nu'}$.
On the intermediate scales of interest, the one-halo term is effectively independent of scale, so is almost entirely degenerate with the shot noise parameters.
Thus we only include the shot noise as a free parameter, as our posterior marginalization absorbs the one-halo terms contribution into the shot noise.
These terms only affect the small scales of the auto-spectra, which provide no appreciable constraining power on \fnl.

\subsection{Instrumental Effects}\label{sec:instrumentaleffectsmodeling}
Finally, there are additional terms that we include to convert the theory predictions outlined above to the measured quantities that we obtain.

Firstly, our `single-frequency'  CIB measurements are sensitive to a finite bandwidth region in frequency space. Instead of integrating over this finite bandwidth region, we calculate our model at the single-frequency \textit{labels} (i.e. 353, 545, and 857 GHz).
To correct for this, we could perform bandpass integration for every calculation of the model, or we could incorporate the finite width of the bandpass by computing multiplicative colour corrections for a fiducial model.
However, we also include multiplicative calibration uncertainty parameters, $f_{\nu}$, that are degenerate with these colour corrections.
Thus, in practice, we simply compute the spectra at the labelled frequencies and marginalize over these multiplicative parameters.

Secondly, the CIB maps  contain both a pixel window function as a result of their $\mathrm{NSIDE}=2048$ HEALPix\footnote{\url{https://healpix.sourceforge.io/}}\footnote{\url{https://github.com/healpy/healpy/}}~\cite{zonca_healpy_2019, gorski_healpix_2005} pixelization, and a $5$ arcminute Gaussian instrumental beam.
These are included as additional $\ell$-dependent factors $p_\ell$ and $b_\ell$ respectively.

Finally, there is a misnormalization of the reconstructed lensing potential as a result of masking that we account for in a Monte Carlo (MC) estimated transfer function, $A_\ell$, discussed in Section~\ref{sec:MCNorm}.

Thus, calibration parameters $f_\nu$, beam $b_\ell$, pixel window function, $p_\ell$, and MC normalization correction $A_\ell$ are introduced such that our final theory spectra are given by
\begin{equation}
    C_{\ell, \mathrm{theory}}^{\nu\kappa} = A_\ell \,b_\ell \,p_\ell \,f_\nu \,C_{\ell}^{\nu\kappa}\label{eq:multiplicative_factors_nk},
\end{equation} 
and
\begin{equation}
    C_{\ell, \mathrm{theory}}^{\nu\nu'} =b_\ell^2p_\ell^2f_\nu f_{\nu'}C_\ell^{\nu\nu'}\label{eq:multiplicative_factors_np}.
\end{equation}

A list of all free parameters can be found in Table~\ref{tab:priors}.

\section{Data}\label{sec:data}

In this section we describe our data products and vector.
We describe the CMB lensing map in Section~\ref{sec:cmblensing}, the CIB maps in Section~\ref{sec:data_cib}, our sky masks in Section~\ref{sec:masks}, and our data vector of cross- and auto-spectra in Section~\ref{sec:data-vector}.
\subsection{CMB Lensing}\label{sec:cmblensing}
We use the reconstructed CMB lensing convergence of Ref.~\cite{carronCMBLensingPlanck2022a}, which was produced from the \Planck CMB PR4 (NPIPE) maps \cite{planckcollaborationPlanckIntermediateResults2020}.
The NPIPE reanalysis includes~$8\%$ more data than the PR3 analysis~\cite{planckcollaborationPlanck2018Results2020a}, and lower levels of noise and systematics across the maps.
The lensing potential is reconstructed using the Generalized Minimum Variance estimator described in Ref.~\cite{maniyar_quadratic_2021}.
These maps show an improvement in signal-to-noise on the lensing power spectrum by up to $\sim20\%$ when compared to the PR3 lensing reconstruction \cite{planckcollaborationPlanck2018Results2020b}.

\subsection{CIB}\label{sec:data_cib}
The CIB is a component of the signal measured in CMB experiments that  dominates over the CMB at high frequencies and is often considered a source of contamination. Many efforts have been made to produce clean maps of the CIB for its own study (e.g.~\cite{penin_accurate_2012, lenzLargescaleMapsCosmic2019, mccarthyLargescaleGalacticdustcleanedCosmic2024}),
and to understand how to remove it to explore other observables (e.g. \cite{madhavacheril_atacama_2020,mccarthy_component-separated_2024,2024PhRvD.109f3530C,2026arXiv260211279M}). In this analysis we consider it as a \textit{signal}.
We use  the~\McCCIB{} maps produced from \Planck NPIPE data~\cite{planckcollaborationPlanckIntermediateResults2020} in the 353, 545, and 857 GHz channels cleaned with neutral Hydrogen data from the HI4PI survey~\cite{hi4picollaboration:HI4PIFullskySurvey2016}.
At 545 and 857 GHz, the CIB dominates over the CMB on all relevant scales.
However, at 353 GHz the CMB dominates over the CIB, and so an estimate of the CMB was directly subtracted from this channel in~\McCCIB.\footnote{See \McCCIB{} for a discussion of why spurious CIB contamination in the CMB estimate is not large enough to oversubtract any CIB signal.}

A commonly used signal-extraction technique in CMB analyses is the Internal Linear Combination~(ILC) ~\cite{1992ApJ...396L...7B,2009A&A...493..835D}, which relies on multi-frequency maps and knowledge of the frequency dependence of the signal of interest.  
Under the assumption that the signal is simply rescaled at each frequency according to its SED, one can compute the optimal weighting of each frequency map to produce a minimum-variance estimate of the desired signal.

It is difficult to separate the CIB and galactic dust through multifrequency ILC methods due to our lack of knowledge of the intrinsic CIB frequency dependence, as well as the very similar SEDs of the CIB and galactic dust.
Previous attempts have been made with a \textit{generalized} ILC~\cite{2011MNRAS.418..467R,2016A&A...596A.109P}, which uses some information about the assumed power spectrum of the CIB; however it is a known issue that the \textit{Planck} generalized needlet ILC (GNILC) CIB map has oversubtracted the CIB on large scales (see, e.g.~\cite{2019A&A...621A..32M}).
Instead, it is helpful to clean the galactic dust with \textit{external} tracers of the dust.
This is often achieved by implementing an ILC-like framework with just a single-frequency CIB measurement and an external HI column density measurement, as the HI column density is highly correlated with the Galactic dust emission~\cite{1996A&A...312..256B}, and purely galactic in origin.
In such a method, knowledge of the CIB SED is not needed as it suffices to assume that there is no emission from the CIB in the HI measurement, and thus that everything in the HI data correlated with the single-frequency measurement can be used to subtract dust.
This approach is commonly taken for CIB power spectrum measurements~\cite{planck_collaboration_planck_2014,lenzLargescaleMapsCosmic2019}; in~\McCCIB{} this was explicitly framed as an ILC-like problem. 

\begin{figure}
    \centering
    \includegraphics[width=\linewidth]{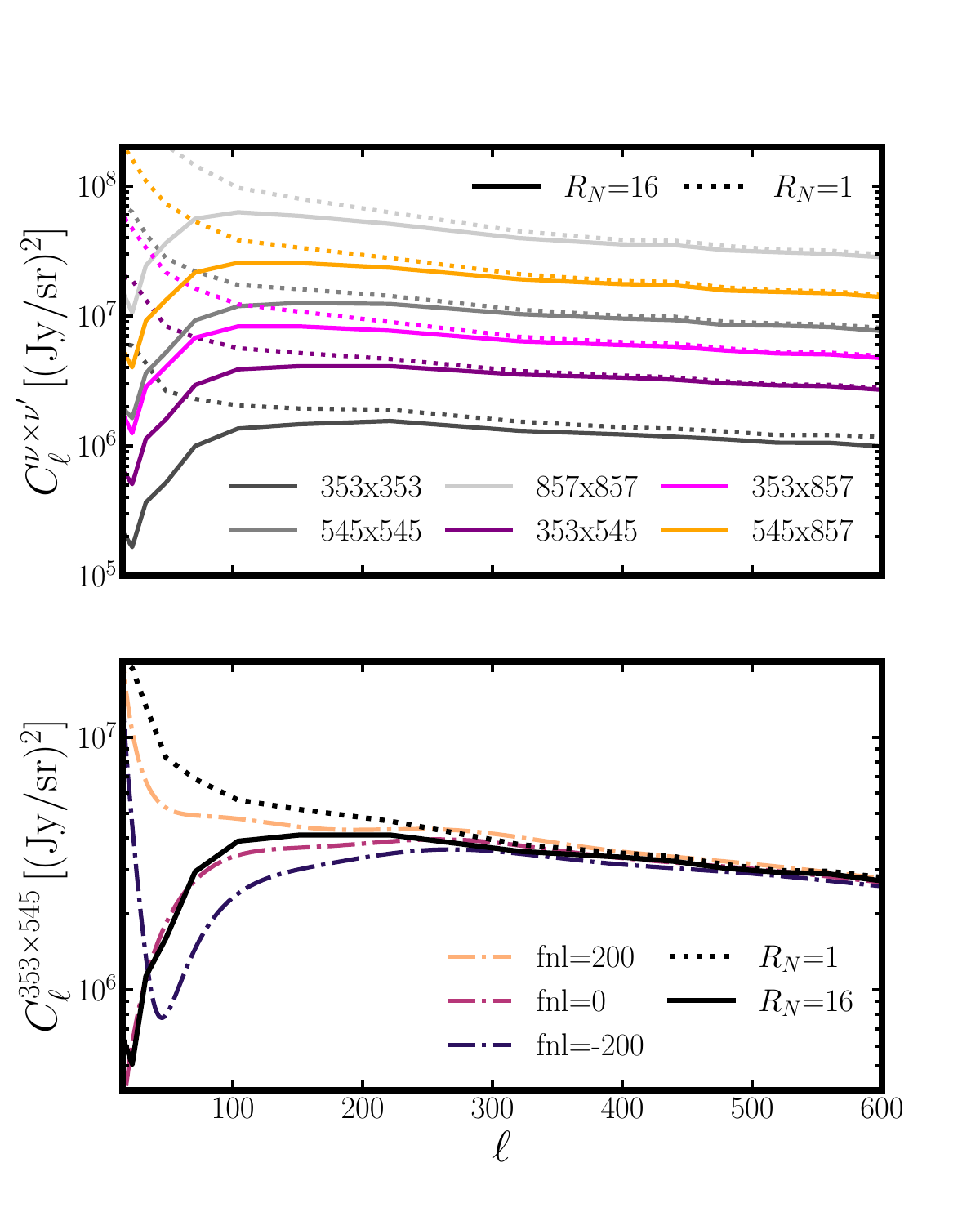}
    \caption{The top panel shows a comparison of the cross- and auto-frequency spectra for the CIB when switching between the unbiased $R_N=1$ maps (dotted) and the biased but cleaner $R_N=16$ maps (solid).
    All curves in the top panel are for spectra on the Cleaner sky region.
    The bottom panel shows the effect of varying \fnl{} on a well fitting model.
    This demonstrates that low-$\ell$ power in the CIB-only cross- and auto- spectra could be misinterpreted as scale-dependent bias, providing erroneous constraining power on \fnl.
    As such, in this analysis, only multipoles in the range $500 \le \ell \le 600$ are used for $C_\ell^{\nu\nu'}$.
    The excess dust contamination in the CIB on large scales also motivates our cross-correlation only analysis with CMB lensing, to mitigate the impact of galactic dust contamination.}
    \label{fig:spectra_comp}
\end{figure}

In~\McCCIB{}, this ILC was performed (using the public ILC code \texttt{pyilc}\footnote{\url{https://github.com/jcolinhill/pyilc}}~\cite{mccarthy_component-separated_2024}) as a spatially-varying real-space ILC that subtracted the mean measured over patches given by a certain angular size. In order to compare with \Lenz{} this angular size was quoted in terms of a HEALPix pixellization scheme; in particular, \Lenz{} subtracted the mean over discrete regions corresponding to NSIDE=16 HEALPix pixels. To avoid confusion with the actual pixellization of our maps, we will refer to this hyperparameter as $R_N$. The choice of $R_N=16$ in \Lenz{} meant that the maps produced were only unbiased at  $\ell \gtrsim 70$.
In \McCCIB{} the large-scale power was retained by cleaning with a larger patch size, corresponding to $R_N=1$ pixelization.
This means there is also additional dust power retained at larger scales, but, importantly, that the \McCCIB{} maps are unbiased to much lower~$\ell$.
{In \McCCIB{} they show that the maps produced are at most $1\%$ biased due to this procedure on all multipoles $\ell\gtrsim10$.}
In this work we use $\ell_\mathrm{min} = 19$.
The differences between these two cleaning patch sizes on the auto and inter-frequency cross power spectra of the CIB are shown in the top panel of Fig.~\ref{fig:spectra_comp}; these are shown for the Cleaner sky {region} {(of two sky regions that we use for analysis; see Section~\ref{sec:masks})}.
For consistency between cleaning algorithms and data products, we exclusively use \McCCIB{} maps throughout this work.
As such, both the $R_N=1$ and $R_N=16$ maps displayed in Fig.~\ref{fig:spectra_comp} were produced in \McCCIB.
The bottom panel of Fig.~\ref{fig:spectra_comp} shows model predictions for the CIB auto spectrum for different values of \fnl, alongside the measured CIB cross-frequency spectra for $353$ and $545$ GHz.
The measured CIB cross-frequency spectra are shown for the maps cleaned with both choices of $R_N$.
One can see that these two maps cover drastically different values of \fnl, and as such, without perfect unbiased removal of the foreground dust in the lower multipoles, $C_\ell^{\nu\nu'}$ cannot be used for \fnl{} constraints.
We therefore only fit an intermediate $\ell$ range of $500 \le \ell \le 600$ for the CIB-only cross- and auto-frequency spectra, to minimize the erroneous constraining power on \fnl{} while breaking necessary degeneracies with the linear bias and redshift evolution parameters.

We assume $C_\ell^{\nu\kappa}$ is unbiased by foreground dust contamination.
Thus we consider all scales on which the CIB maps are unbiased to signal oversubtraction, corresponding to a large-scale minimum multipole cut of $\ell_\mathrm{min}=19$ in our analysis.
As \Planck lensing reconstruction is derived mostly from high-$\ell$ temperature fluctuations, the leading order contribution from galactic dust (D) in the cross-correlation with the CIB corresponds to a $\langle D_\mathrm{low-\ell}D_\mathrm{high-\ell}D_\mathrm{high-\ell}\rangle$ bispectrum, with two high-$\ell$ legs, $D_\mathrm{high-\ell}$, from the lensing quadratic estimator, and one leg, $D_\mathrm{low-\ell}$, from the low-$\ell$ contamination of the CIB.
As dust power is small in the high-$\ell$s used in lensing reconstruction, and as the PR4 lensing reconstruction uses SMICA \cite{cardosoComponentSeparationFlexible2008} foreground-cleaned CMB maps, we expect a negligible bias from this bispectrum.
We test this in Section~\ref{sec:dust_bias} by comparing constraints on two independent sky regions.

\begin{figure}
    \centering
    \includegraphics[width=\linewidth]{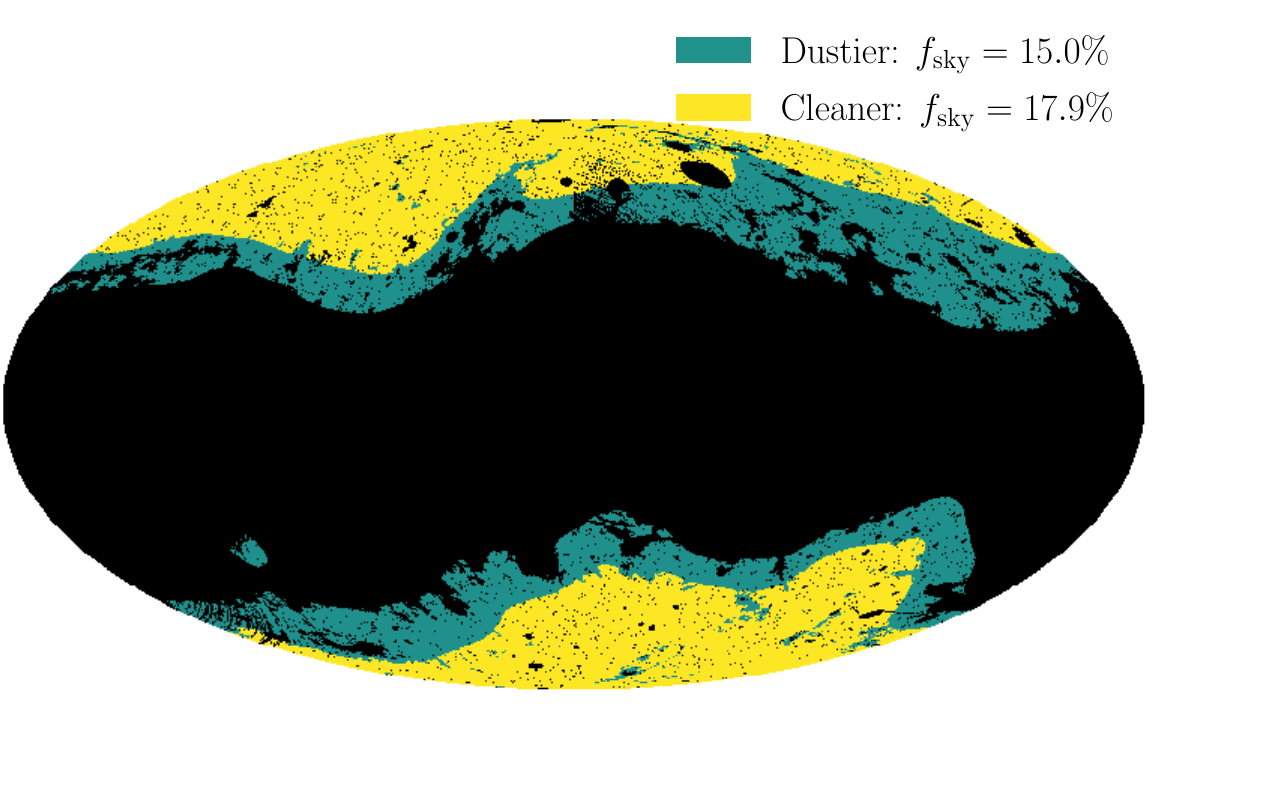}
    \caption{Masks of the two regions considered for analysis. The Dustier region surrounds the Cleaner region.
    The sky is shown in Mollweide projection in galactic coordinates, with the Galactic plane as the central horizontal line of the ellipse.
    We can see that the Dustier region is closer to the galactic plane.
    }
    \label{fig:doughnut_vis}
\end{figure}

\subsection{Masks}\label{sec:masks}

Throughout this work we use the \pypackage{pymaster}\footnote{\url{https://github.com/LSSTDESC/NaMaster}} package~\cite{alonsoUnifiedPseudoFramework2019}, a python implementation of the \software{master} algorithm \cite{hivon_master_2002}, to remove mask-induced mode coupling.
We consider sky regions defined by the level of foreground dust contamination, and we use masks produced in \Lenz{} as a proxy for this.
As an overview these masks are the intersection between the \Planck galactic plane mask, and a boolean mask defined by regions of the HI4PI Survey \cite{hi4picollaboration:HI4PIFullskySurvey2016} below a given neutral hydrogen column density, $n_\mathrm{HI}$.
For a more detailed description see \Lenz.

\begin{figure}
    \centering
    \includegraphics[width=\linewidth]{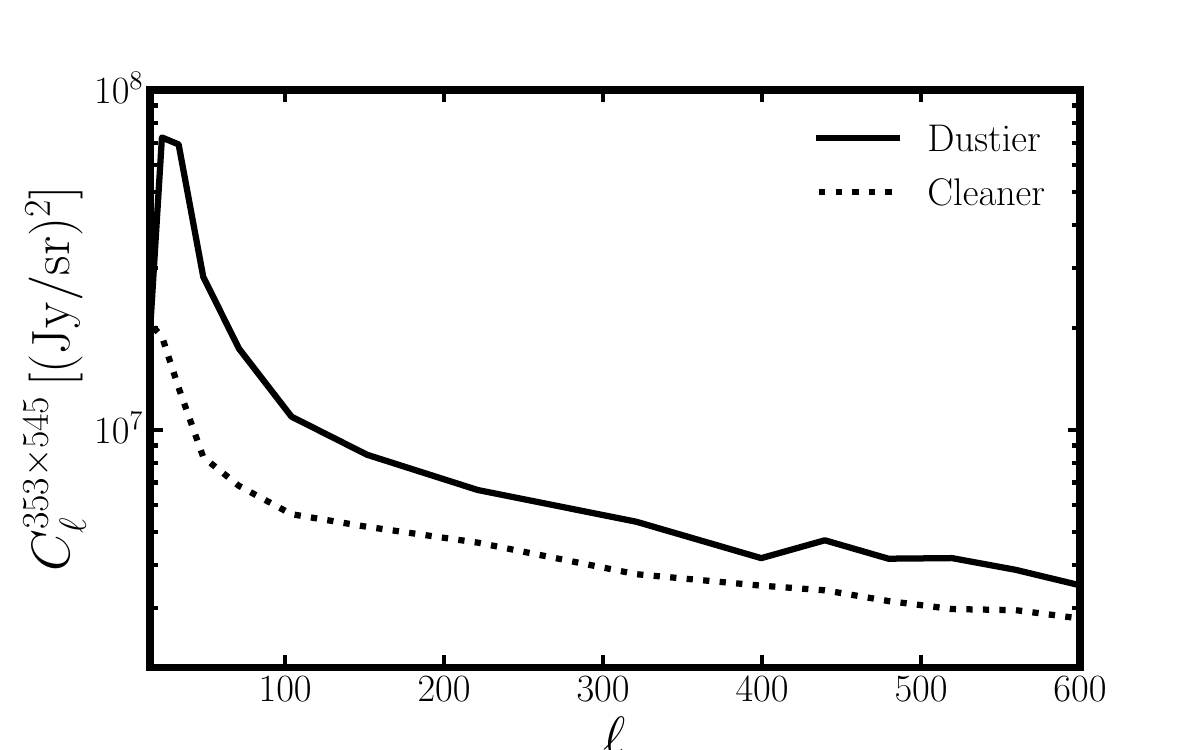}
    \caption{This figure compares the same cross-frequency spectra for the CIB in both the Cleaner and Dustier region, demonstrating the residual dust power remaining in the Dustier region.
    In both cases the $R_N=1$ maps are used.}
    \label{fig:spectra_comp_dust_level}
\end{figure}

We utilise two \Lenz{} masks in construction of two independent sky regions.
The first \Lenz{} mask is defined by the overlap of the HI4PI Survey sky where~$n_\mathrm{HI}<2.5\times 10^{20}\,\mathrm{cm}^{-2}$ and the $20\%$ galactic plane mask.
The second is defined by the overlap of the HI4PI Survey sky where $n_\mathrm{HI}<4.0\times 10^{20}\,\mathrm{cm}^{-2}$ and the $40\%$ galactic plane mask, which therefore contains the first region.
\Lenz{} produced one of these masks per frequency band.
However, when using the~\Lenz{} masks in this work, we instead consider a single mask produced from the product of each of the individual frequency Boolean masks.
We then construct two non-overlapping regions, which we call a ``Cleaner'' and a ``Dustier'' sky region from these two masks as follows. 
Our Cleaner region is the~$n_\mathrm{HI}<2.5\times 10^{20}\mathrm{cm}^{-2}$~\Lenz{} region outlined above, while the Dustier region is defined by the complement of the Cleaner region within the $n_\mathrm{HI}<4.0\times 10^{20}\mathrm{cm}^{-2}$ \Lenz{} region.
This produces two independent regions, as shown in Fig.~\ref{fig:doughnut_vis}, with the Dustier region being an annulus around the Cleaner region.
When measuring spectra we use an apodized mask of these two independent sky regions, apodized using \pypackage{pymaster} with the C1 apodization method and an apodization scale of 15'.
In Fig.~\ref{fig:spectra_comp_dust_level} we show the difference in the auto power of the CIB between the two regions.
We can see that the Dustier region has more power than the Cleaner region in $C_\ell^{\nu\nu'}$ across all scales; this is not unexpected, as the Dustier region has significantly more dust contamination.
The sky fraction of the Cleaner region is $f_\mathrm{sky}=17.9\%$, and of the Dustier region is $f_\mathrm{sky}=15.0\%$.

\subsection{Data Vector}\label{sec:data-vector}

We measure our auto- and cross-power spectra with \texttt{pymaster}.
Most of the information on \fnl{} is derived from the larger scales of $C_\ell^{\nu\kappa}$, whereas the CIB parameters are informed by the entirety of $C_\ell^{\nu\kappa}$ along with $C_\ell^{\nu\nu'}$.
We include $C_\ell^{\nu\nu'}$ only on intermediate scales where we are less likely to be biased by remaining dust contamination, while remaining in a regime where we can use our linear model.
While the values of the other parameters are not of interest in this work, constraining them with the extra information provided by $C_\ell^{\nu\nu'}$ will help with degeneracy breaking between these parameters and \fnl.
To tackle both of these goals we use a combination of logarithmically spaced binning for lower $\ell$ and and linearly spaced binning for higher $\ell$ (in particular, at $\ell\ge380$ we use a spacing $\Delta\ell=40$).
The bins are defined by the edges:
\begin{align*}
    \ell_\mathrm{edge} = &19, 27, 40, 58, 85, 124, 180, 263,\\
    &380, 420, 460, 500, 540, 580, 620, \numberthis\label{eq:bin_edges}
\end{align*}
{with each bin containing multipoles including the lower edge, and excluding the upper (e.g the lowest bin is defined by $19\le\ell<27$).}
Our final data-vector is structured as
\begin{equation}
    C_\mathrm{data} = \begin{cases}
        C_\ell^{\nu \kappa, \mathrm{Cleaner}}\\
        C_\ell^{\nu \kappa, \mathrm{Dustier}}\\
        C_\ell^{\nu \nu', \mathrm{Cleaner}}
    \end{cases}\label{eq:data_vector}
\end{equation}
where $\nu, \nu' \in[353, 545, 857]$.
For $C_\ell^{\nu\kappa}$ the full binning is used, whereas for $C_\ell^{\nu\nu'}$ only the final three bins are used.
To remove the impact of instrumental noise bias, measurements of $C_\ell^{\nu\nu'}$ for $\nu = \nu'$ use two independent half-mission splits of the data.
These half-mission splits each contain an independent realisation of the instrumental noise, so their cross-spectrum does not contain any noise power.
For $C_\ell^{\nu\nu'}$ we only use the Cleaner region in our data vector, as we see residual dust power remaining in the~$500\le\ell\le600$ range for the Dustier region, as shown  in Fig.~\ref{fig:spectra_comp_dust_level}. The Cleaner region is assumed to have no residual dust at high $\ell$, as the CIB auto-spectra for these scales do not change when measured on smaller but less dusty regions of the sky.

\section{Likelihood}\label{sec:likelihood}
We produce our posterior distributions by exploration of the parameter space through Markov chain Monte Carlo (MCMC) sampling, within the framework of \pypackage{cobaya}\footnote{\url{https://github.com/CobayaSampler/cobaya}}~\cite{torrado_cobaya_2021, noauthor_asclnet_nodate}.
Constraints on \fnl{} are then obtained by marginalization over all other model parameters.
We use a Gaussian likelihood $\mathcal{L}$, such that
\begin{equation}
\begin{split}
    -2\ln\mathcal{L} \propto &\left[C_\mathrm{data} - C_\mathrm{theory}(\Theta)\right]^\intercal \mathbb{C}^{-1}\left[C_\mathrm{data} - C_\mathrm{theory}(\Theta)\right]
    \\ &+ \chi^2_\mathrm{priors}(\Theta)\label{eq:likelihood},
\end{split}
\end{equation}
where $\mathbb{C}$ is our covariance matrix which we will describe in Section~\ref{sec:covariance}; $C_\mathrm{data}$ is the data vector defined in Equation~\eqref{eq:data_vector}; $C_\mathrm{theory}$ is a theory vector calculated from Equations~\eqref{eq:multiplicative_factors_nk} and \eqref{eq:multiplicative_factors_np} for a set of model parameters $\Theta$; and $\chi^2_{\mathrm{priors}}(\Theta)$ quantifies our priors, which will be described in Section~\ref{sec:priors}.
A list of all model parameters is given in Table~\ref{tab:priors}.
{Finally, Section~\ref{sec:MCNorm} describes the estimation of the MC normalisation correction, $A_\ell$, mentioned in Section~\ref{sec:instrumentaleffectsmodeling}.}

\subsection{Covariance matrix}\label{sec:covariance}

We estimate the covariance matrix $\hat{\mathbb{C}}$ from 480 Gaussian simulations, designed to reproduce the relevant power on a given sky region and including half-mission splits for the CIB auto-spectra.
The simulations are masked identically to the data to ensure masking effects are captured.
They match the observed CIB power and include CIB-$\kappa$ correlations consistent with a fiducial model, derived from maximum-likelihood parameters obtained using an analytic covariance built from the fiducial values in Table~\ref{tab:priors}.

For details, see Appendix~\ref{app:sims}; a brief summary follows.

We build the simulations from the NPIPE lensing simulations of Ref.~\cite{carronCMBLensingPlanck2022a}.
The NPIPE simulations provide input $\phi$ maps and corresponding reconstructed $\hat{\kappa}$ maps.
Since our model uses $C_{\ell}^{\nu \kappa}$, we convert $\phi$ to $\kappa$ and generate full-sky CIB realisations with both inter-frequency and $\kappa$ correlations, such that we match the fiducial $C_\ell^{\nu\kappa}$ theory.
We then add Gaussian dust realisations, which include inter-frequency correlations but are uncorrelated with the CIB.
The dust power for these Gaussian realisations is chosen such that the final simulations (CIB + dust) reproduce the measured $C_\ell^{\nu\nu'}$.

\begin{figure}
    \centering
    \includegraphics[width=\linewidth]{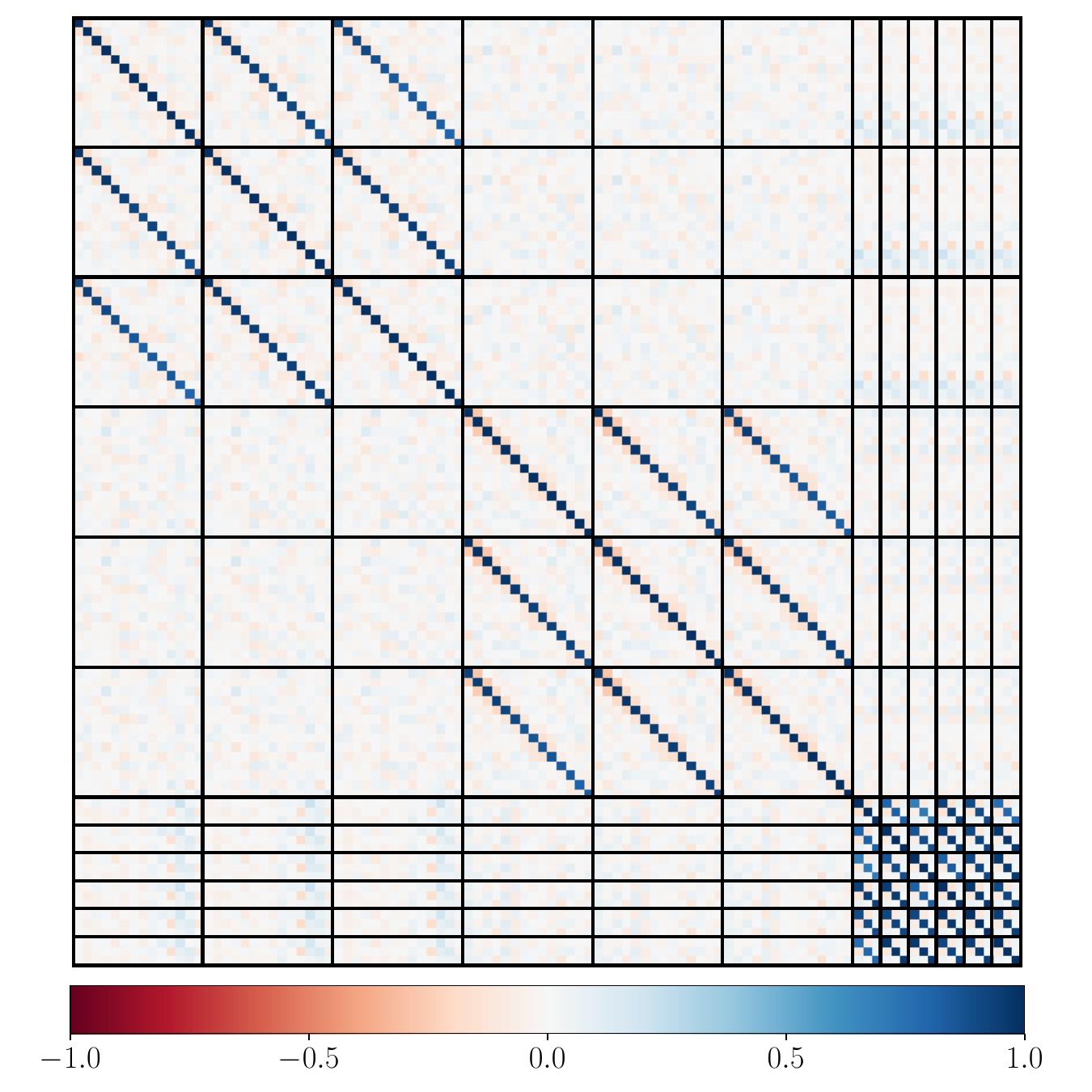}
    \caption{Correlation matrix obtained for 480 Gaussian simulations built on top of the lensing simulations of \cite{carronCMBLensingPlanck2022a}.
    The order of black outlined sections from left to right are: Cleaner [$C_\ell^{353\kappa}$, $C_\ell^{545\kappa}$, $C_\ell^{857\kappa}$], Dustier [$C_\ell^{353\kappa}$, $C_\ell^{545\kappa}$, $C_\ell^{857\kappa}$], Cleaner [$C_\ell^{353,353}$, $C_\ell^{353,545}$, $C_\ell^{353,857}$, $C_\ell^{545,545}$, $C_\ell^{545,857}$, $C_\ell^{857,857}$].
    }
    \label{fig:corr_mat}
\end{figure}

\begin{figure}
    \centering
    \includegraphics[width=\linewidth]{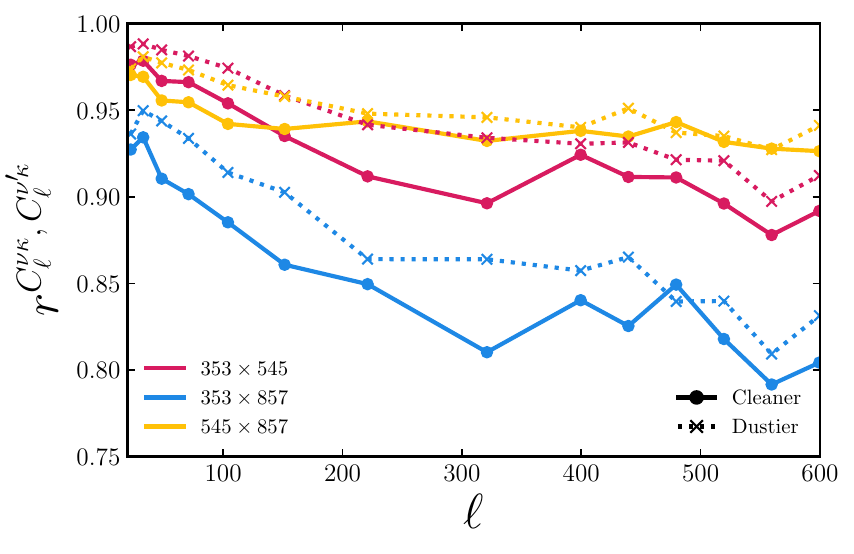}
    \caption{The correlation coefficients, $r^{C_\ell^{\nu\kappa},C_\ell^{\nu'\kappa}}$ between different cross-spectra, $C_\ell^{\nu\kappa}$, at the same multipole.
    We see there is a high degree of correlation between all frequencies in a given sky region (Cleaner or Dustier).
    The correlation coefficients for the Cleaner-Dustier correlations are omitted as these are all negligible.}
    \label{fig:corr_plot}
\end{figure}

\begin{figure}
    \centering
    \includegraphics[width=\linewidth]{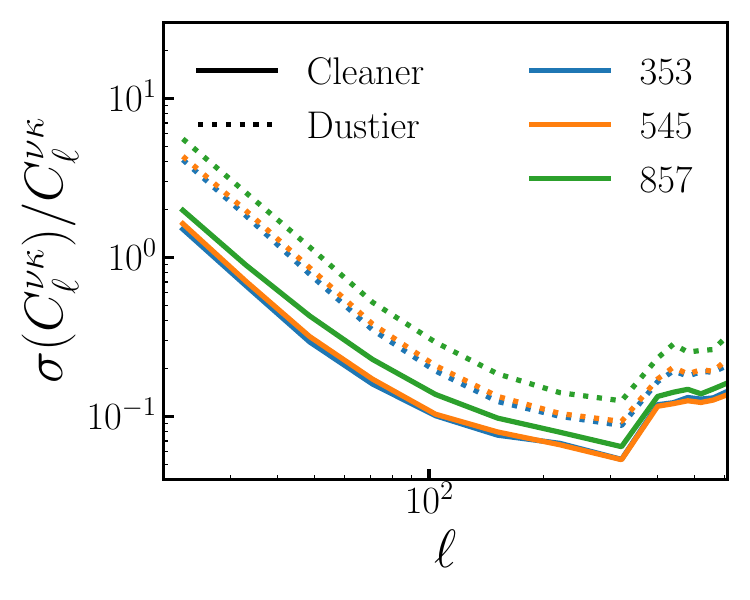}
    \caption{Fractional errors on $C_\ell^{\nu\kappa}$ estimated from the diagonal of the covariances matrix we estimated from simulations,
    with $C_\ell^{\nu\kappa}$ being the input for the simulations.}
    \label{fig:fractional_error}
\end{figure}

As $\hat{\mathbb{C}}$ is estimated from a finite number of simulations, we apply the Hartlap correction \cite{hartlapWhyYourModel2007} to ensure that $\mathbb{C}^{-1}$ is unbiased.
This unbiased estimate is
\begin{equation}
    \mathbb{C}^{-1} = \frac{n-p-2}{n-1} \hat{\mathbb{C}}^{-1}\label{eq:hartlap},
\end{equation}
where $n$ is the number of simulations in our estimate and $p$ is the size of our data vector.
The final correlation matrix is displayed in Fig.~\ref{fig:corr_mat}.
The level of correlations between different cross-spectra at the same multipole is more easily seen in Fig.~\ref{fig:corr_plot}.
We can see that there is a high degree of correlation between $C_\ell^{\nu\kappa}$ for different frequencies within the same sky region {and multipole}; however, the Dustier and Cleaner regions are uncorrelated as one would expect for non-overlapping regions.
We also see a large amount of correlation between all $C_\ell^{\nu\nu'}$.
In Fig.~\ref{fig:fractional_error} we show the fractional uncertainty in $C_\ell^{\nu\kappa}$, taken from the diagonal of our estimated covariance matrix (square rooted), and scaled by a fiducial theory data vector.\footnote{This is the same fiducial theory used to generate the covariance matrix.}
We can see that in both the Cleaner and Dustier regions the fractional uncertainty increases with frequency.
This result is consistent with the SED of the galactic dust rising faster with frequency than CIB SED.
We can also clearly see that the Dustier region has a visually larger fractional uncertainty than the Cleaner region.

\subsection{Priors}\label{sec:priors}
\begin{table}[t]
    \centering
    \begin{tabular}{c c c}
    \toprule
         Parameter & Fiducial & Prior \\
         \toprule
         $f_\mathrm{NL}$ & 0 & $\mathcal{U}(-1000, 1000)$ \\
         \hline
         $b_0$ & 0.83 & $\mathcal{N}(0.83, 0.11)$ \\
         $b_1$ & 0.742 & $\mathcal{U}(0, 200)$ \\
         $b_2$ & 0.318 & $\mathcal{U}(0, 200)$\\
         $\log_{10}\alpha$ & -2.155 & $\mathcal{U}(-100, 100)$ \\
         $\beta$ & 3.590 & $\mathcal{U}(-10, 10)$ \\
         $\log_{10}\gamma$ & 0.390 & $\mathcal{U}(-50, 50)$ \\
         $\delta$ & 6.578 & $\mathcal{U}(-10, 10)$ \\
         \hline
         $S_{353, 353}$ & 197 & $\mathcal{U}(171.0, 279.0)$ \\
         $S_{353, 545}$ & 504 & $\mathcal{U}(412.2, 673.8)$ \\
         $S_{353, 857}$ & 839 & $\mathcal{U}(693.4, 1132.6)$ \\
         $S_{545, 545}$ & 1393 & $\mathcal{U}(1104.8, 1803.2)$ \\
         $S_{545, 857}$ & 2461 & $\mathcal{U}(2017.8, 3292.2)$ \\
         $S_{857, 857}$ & 5186 & $\mathcal{U}(4276.8, 6979.2)$ \\
         \hline
         $f_{353}$ & 1.0 & $\mathcal{N}(1, 0.05)$\\
         $f_{545}$ & 1.0 & $\mathcal{N}(1, 0.05)$\\
         $f_{857}$ & 1.0 & $\mathcal{N}(1, 0.05)$\\
    \end{tabular}
    \caption{Overview of the priors on the parameter space used in our analysis.
    These are our PNG parameter,~\fnl; our SFRD parameters, $\alpha, \beta, \gamma, \delta$; our bias parameters,~$b_0, b_1, b_2$; our shotnoise parameters $S_{\nu,\nu'}$; and our calibration parameters, $f_\nu$.
    The priors are from either Gaussian distributions with mean $\bar{x}$ and standard deviation~$\sigma$, denoted by $\mathcal{N}(\bar{x}, \sigma)$; or Uniform distributions with range~$[x, y]$, denoted by $\mathcal{U}(x, y)$.
    }
    \label{tab:priors}
\end{table}

\begin{table}
    \begin{tabular}{c c}
    \toprule
         $\nu$ & $\nu \bar{I}_\nu$ [$\mathrm{nW m}^{-2}\mathrm{Sr}^{-1}$] \\
         \toprule
         $3000$ & $12.61\substack{+8.31\\-1.74 }$\\
         $1875$ & $13.63\substack{+3.53\\-0.85 }$\\
         $1200$ & $10.2\substack{+2.6\\-2.3 }$\\
         $857$ & $6.6\substack{+1.7\\-1.6 }$\\
         $600$ & $2.8\substack{+0.93\\-0.81 }$\\
         $353$ & $0.46\substack{+0.04\\-0.05 }$\\
    \end{tabular}
    \caption{The priors on CIB intensities, collated in Ref.~\cite{maniyarHistoryStarFormation2018a} and based on data from Refs.~\cite{berta_building_2011}, \cite{berta_building_2011}, \cite{bethermin_unified_2012}, \cite{bethermin_unified_2012}, \cite{bethermin_unified_2012}, and \cite{zavalaSCUBA2CosmologyLegacy2017} respectively.}
    \label{tab:means}
\end{table}
The set of priors and fiducial values of the model parameters is given in Table~\ref{tab:priors}, with fiducial values taken from the best-fit values of Ref.~\cite{maniyarHistoryStarFormation2018a}.
We have some additional prior information about the CIB bias, calibration parameters, and the shot noise, as described below.
Except for these, all priors are uninformative.

\subsubsection{Bias priors}
There are two priors on the CIB bias parameters.
First we impose a present day prior such that $b_0 = 0.83\pm0.11$, taken from Ref.~\cite{maniyarHistoryStarFormation2018a}, which was converted from the constraints on $b_I\sigma_8$ from Ref.~\cite{saunders_spatial_1992}, using $\sigma_8$ measurements of Ref.~\cite{planck_collaboration_planck_2016}.
The second prior is a physicality prior that enforces the linear Gaussian bias to be positive for all $z$.
\subsubsection{CIB mean}
A calculation of the CIB mean frequency weighted intensity, $\nu I_\nu$, from Equation~\eqref{eq:CIB_intensity}, is imposed as an additional prior, with each frequency having a split normal distributon. These priors are listed in Table~\ref{tab:means}, and were calculated from a range of different galaxy number count measurements \citep{berta_building_2011, bethermin_unified_2012, zavalaSCUBA2CosmologyLegacy2017}, and collated in Ref.~\cite{maniyarHistoryStarFormation2018a}.

\subsubsection{Shot noise and 1-halo term}
We follow a similar procedure to Ref.~\cite{planck_collaboration_planck_2014}, which uses a flat prior in the $1\sigma$ region of the shot noise calculated from the model of Ref.~\cite{bethermin_unified_2012}. 
As the one-halo term is highly degenerate with the shot noise, and the fiducial one-halo term is $\lesssim20\%$ the size of~$S_{\nu, \nu'}$ throughout the regime of interest, we incorporate uncertainty on the one-halo term by broadening the width of the shot noise prior by 20\% (and never varying the one-halo term).

\subsection{MC normalization correction}\label{sec:MCNorm}

As mentioned in Section~\ref{sec:cib_model}, reconstructed lensing maps produced by a quadratic estimator on the masked CMB are slightly misnormalized due to the masking~\cite{benoit-levy_full-sky_2013}.
As such, we also use our simulations to calculate the relevant correction to this misnormalization.
We can produce an MC estimate of the transfer function of this misnormalization for our specific masked case by comparing the cross-spectra of our simulations with the reconstucted $\kappa$ simulations.
This factor is given by
\begin{equation}
    A_\ell = \frac{\langle \hat{C}_\ell^{\nu_{\mathrm{mask}} \hat{\kappa}_{\mathrm{mask}}} \rangle}{C_{\ell, \mathrm{theory}}^{\nu\kappa}}\label{eq:MCnorm},
\end{equation}
where $\hat{C}_\ell$ is the measured angular power spectra of the reconstructed version, $\hat\kappa$, of the input simulation,~$\kappa$, used to produce the correlated CIB field $\nu$, and~$\langle\cdot\rangle$ is the average over the set of simulations.
{The $C_{\ell, \mathrm{theory}}^{\nu\kappa}$ in the denominator is the input theory used to generate the simulations.}
The correction, $A_\ell$, is then applied to our theory $C_\ell^{\nu\kappa}$ when used in our likelihood.
See Appendix~\ref{app:sims} for the estimated transfer functions on the Cleaner and Dustier regions.

\section{Results}\label{sec:results}

In this Section we present our results and various consistency tests.
We present the results of our fiducial analysis in Section\ref{sec:fid_analysis}.
We explore our stability to the bias parametrization in Section~\ref{sec:alternative_bias}.
We discuss our sensitivity to any potential galactic dust bias in Section~\ref{sec:dust_bias}.
Finally, we discuss the frequency dependence of our results in Section~\ref{sec:frequency_dependence}.

\subsection{Fiducial Analysis}\label{sec:fid_analysis}
\begin{figure}
    \centering
    \includegraphics[width=\linewidth]{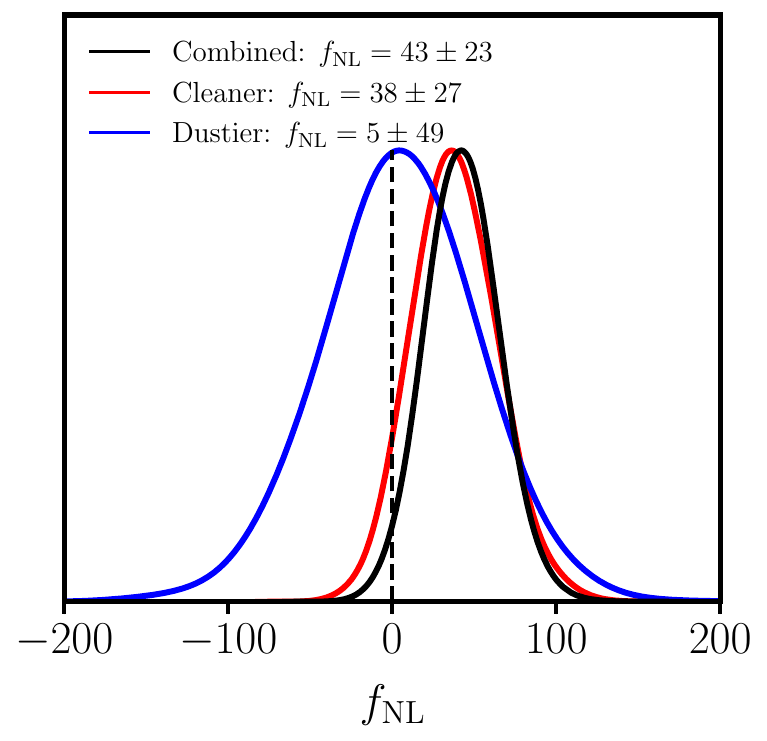}
    \caption{Posterior on local-type PNG parameter \fnl, for the combined analysis (black), Cleaner (red) and Dustier (blue) regions.
    The dashed line indicates $f_\mathrm{NL}=0$.
    We see consistency between the Cleaner and Dustier posteriors, with the Cleaner region dominating the combined constraints.
    We find results consistent with $f_\mathrm{NL}=0$ within $2\sigma$.
    {The full 2-dimensional posterior distributions for these cases is shown in Appendix~\ref{app:corner_main}.}
    }
    \label{fig:posterior_fnl}
\end{figure}

The final 1-dimensional posterior on \fnl{} for the combined analysis is shown in Fig.~\ref{fig:posterior_fnl}, giving $f_\mathrm{NL}= 43 \pm 23$. This is consistent with zero at $2\sigma$.

The Cleaner region contributes the majority of the constraining power ($\sigma(f_\mathrm{NL}) = 27$), with a modest $\sim15\%$ improvement in $\sigma(f_\mathrm{NL})$ upon inclusion of the Dustier region.
Comparing the two regions, the Cleaner and Dustier constraints are consistent at the $1\sigma$ level; this is shown in Fig.~\ref{fig:posterior_fnl}.
Compared to~\McCfNL, which found $\sigma(f_\mathrm{NL})=40$ for their most constraining region, we find an overall factor $\sim 2$ improvement in $\sigma(f_\mathrm{NL})$.
This is driven mostly by new maps that are unbiased to larger angular scales, as well as a more optimal treatment of additional sky area.
\begin{figure*}
    \centering
    \includegraphics[width=\linewidth]{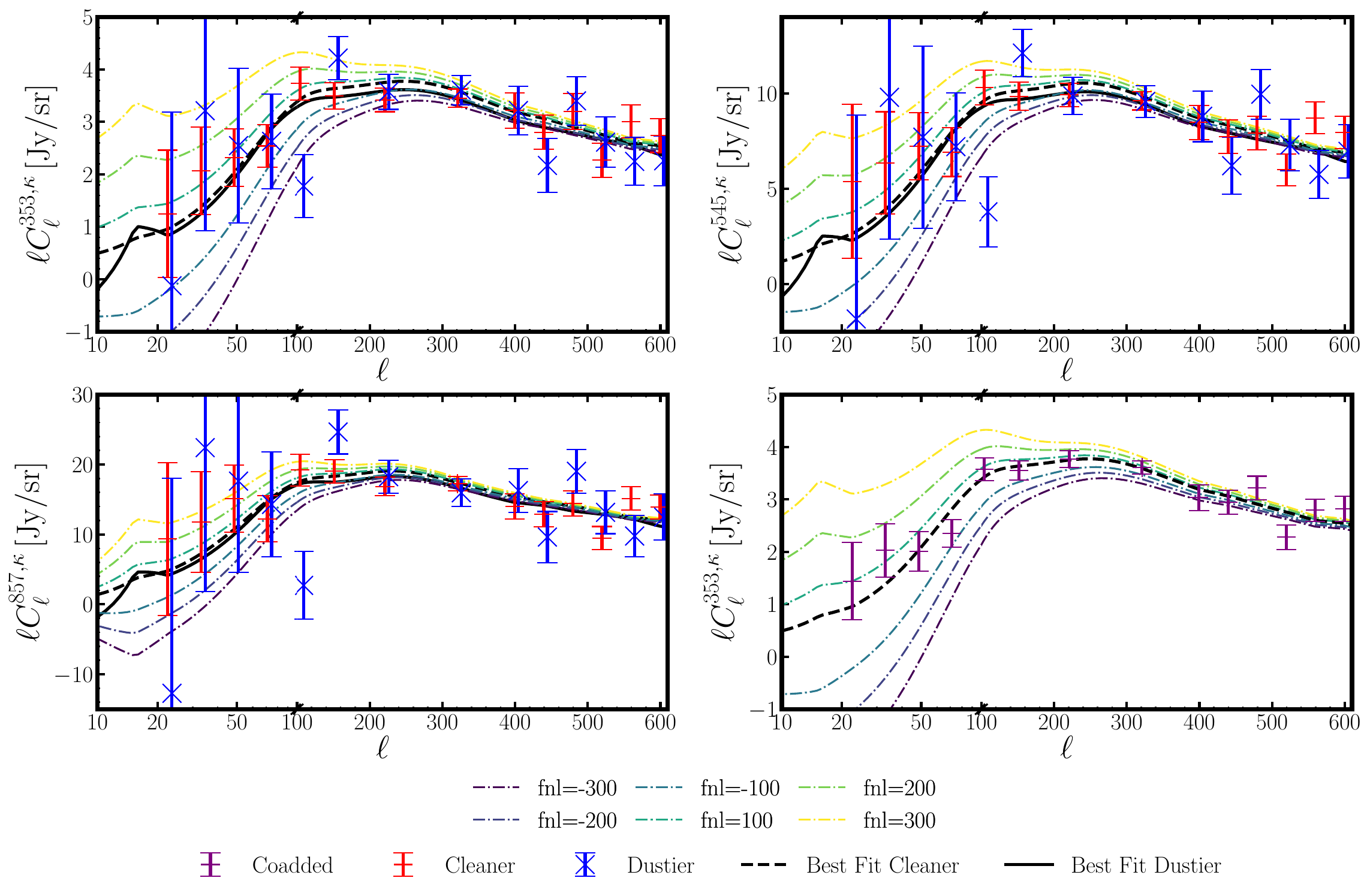}
    \caption{
    Comparison of best-fitting model with variations in \fnl, along side measurements of the CIB$\times$CMB lensing cross-spectra. Error bars are estimated from simulations (See Section~\ref{sec:covariance} and Appendix~\ref{app:sims}).
    {The best fitting theory curves are shown in black, with the dashed curves being produced with the Cleaner region's transfer function, and the solid with the Dustier region's (see Section~\ref{sec:MCNorm} and Appendix~\ref{app:sims}).}
    Measured spectra for both the Cleaner and Dustier regions are shown.
    The bottom right panel shows the same theory curves as the top left, but now we have coadded all of our data on to the $353$ GHz channel for a visual representation of the combined constraining power of the data.
    See Section~\ref{sec:fid_analysis} for details of the coadding procedure.
    }
    \label{fig:modelcomparissons}
\end{figure*}
Fig.~\ref{fig:modelcomparissons} shows the data points along with a range of theory curves fixed to the maximum a posteriori (MAP) parameters, but with varying \fnl.
As the different frequencies are highly correlated it is difficult to interpret the combined constraining power of the data.
To help visualize this, the lower right panel of Fig.~\ref{fig:modelcomparissons} displays a coadded version of all our data (two sky regions and three frequencies).
We coadd the data as follows. First, we inverse scale each frequency's cross-spectra with $\kappa$ by the best-fitting model, $C_\ell^{\nu\kappa}$, so that each is effectively an estimate of unity.
We then produce an inverse variance weighted mean of these spectra using our full covariance matrix (also rescaled).
Finally this new estimate is rescaled by the best-fitting model $C_\ell^{\nu\kappa}$ for the 353 GHz channel for visualization.

Despite high inter-frequency correlations within a given sky region, the Cleaner and Dustier regions probe separate sky areas and are thus largely independent.
This is easily seen by comparing the off-diagonal $3\times 3$ blocks of Fig.~\ref{fig:corr_mat}.
We note that the central value of the \fnl{} posterior for the combined region is slightly higher than both the Cleaner and Dustier regions separately.
This is likely as a result of degeneracy breaking in the high-dimensional parameter space.
The 2-dimensional posterior distributions for these cases are shown in Appendix~\ref{app:corner_main}.

\begin{figure}
    \centering
    \includegraphics[width=\linewidth]{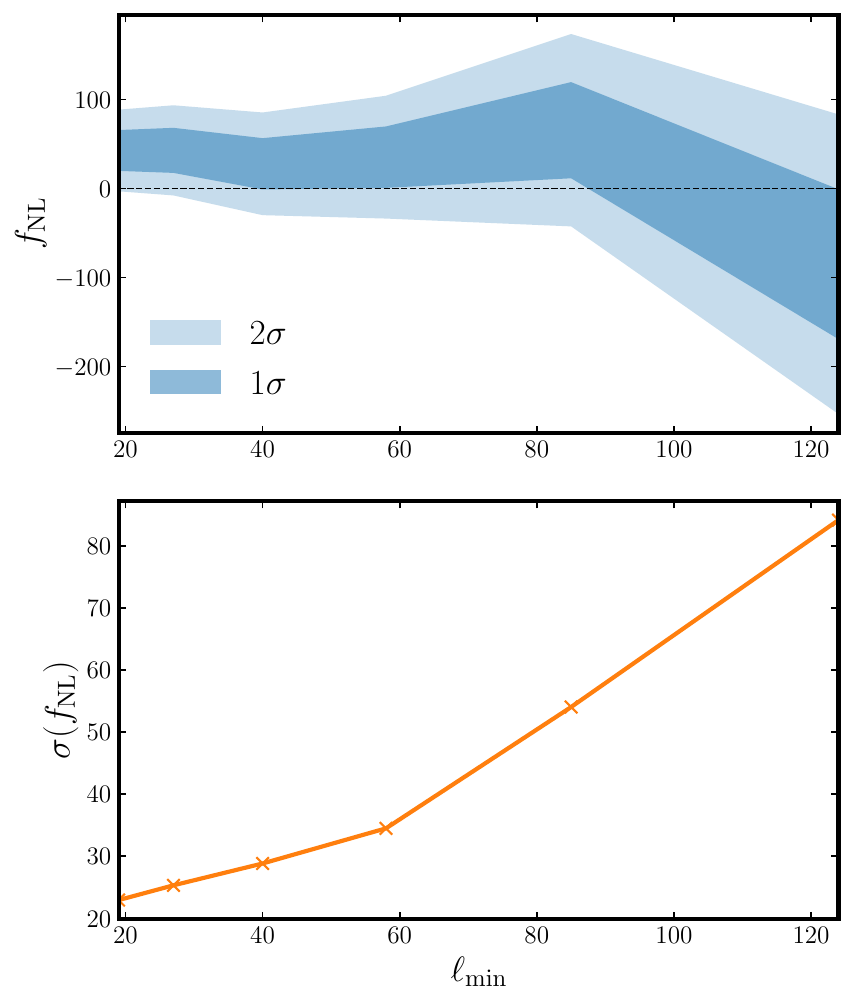}
    \caption{{Constraints on \fnl{} when varying the minimum multipole, $\ell_\mathrm{min}$ used in our analysis.
    The top panel shows the $1\sigma$ and $2\sigma$ intervals for \fnl{} as $\ell_\mathrm{min}$ is increased.
    The bottom panel plots the posterior width,~$\sigma(f_\mathrm{NL})$, against the minimum multipole used.
    The constraints on \fnl{} are consistent as $\ell_\mathrm{min}$ is varied, and posterior widths are noticeably reduced by a lower $\ell_\mathrm{min}$.
    }}
    \label{fig:lmin_scaling}
\end{figure}

We can  see the importance of including the lower $\ell$ that the new \McCCIB{} maps provide by considering the \fnl{} constraints changing $\ell_\mathrm{min}$.
This is shown in Fig.~\ref{fig:lmin_scaling}, where we also see that the constraints remain consistent with different choices of~$\ell_{\mathrm{min}}$.

\subsection{Alternative bias parameterizations}\label{sec:alternative_bias}

\begin{figure}
    \centering
    \includegraphics[width=\linewidth]{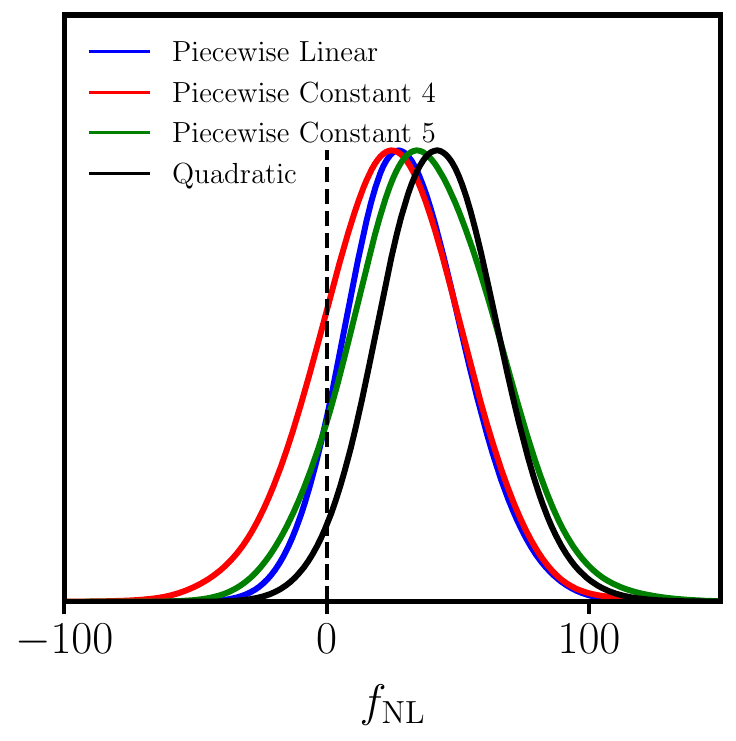}
    \caption{Comparison of the full posteriors of \fnl{} for the different bias parameterization.
    The black indicates the posteriors for our main model, the quadratic bias.
    The piecewise constant parameterizations are separated by the number of bins used.
    We can see that the results and corresponding constraining power are not strongly dependent on bias parameterization.
    }
    \label{fig:alternate_fnl_post}
\end{figure}

As the scale-dependent bias effect has dependence on the Gaussian bias through
\begin{equation}
    \Delta b(z) \propto f_\mathrm{NL}(b^G(z) -1),
\end{equation}
we wish to ensure that our constraints are robust to changes in the parameterization of the bias' redshift evolution.
Following Ref.~\cite{maniyarHistoryStarFormation2018a} and \McCfNL, we have thus far assumed a quadratic evolution given in Equation~\eqref{eq:bCIB}.
As peak star-formation occurs around $z\approx 2$, we expect most information about the bias to be found there.
There could, therefore, be a risk of over-specifying the bias at high redshifts due to enforcement of a quadratic shape throughout, when sensitivity is only present in a small redshift range.
This could result in overconfident constraints on \fnl, by enforcing a specific bias evolution with redshift.
To explore this effect we consider alternative bias parameterizations that are less restrictive on the overall evolution.
In all parametrizations we still impose the Gaussian prior that we previously placed on $b_0$ on the CIB bias at $z=0$.
We explore a piecewise linear and two piecewise constant models, described below.
The resulting posteriors on \fnl in every case are shown in Fig.~\ref{fig:alternate_fnl_post}.

\textbf{Piecewise Linear}: In this model the bias is specified at $z\in [0, 1, 4, 7]$, with linear interpolation between.
We recover the same posterior width as our fiducial quadratic bias case, but with a slightly lower central value.
The results are consistent at the~$\sim0.5\sigma$ level. 

\textbf{Piecewise Constant}: Here the bias is specified as a constant in regions between $[0, 1, 2.5, 4, 7]$ for the 4-bin case, and $[0, 1, 2, 3, 4, 7]$ for the 5-bin case.
Both cases produce similar posterior widths to our fiducial, and the less restrictive, 5-bin, case sees only~$\sim 0.1\sigma$ shift in central value from our fiducial case.

We can see that changing the bias parameterization has a small effect on the central value and width of the posteriors. We conclude that the alternate biases are consistent with our fiducial model, and that our estimate on the posterior width of \fnl{} is not sensitive to bias parameterization.

\subsection{Dust bias}\label{sec:dust_bias}

As mentioned in Section~\ref{sec:data_cib}, there is the possibility of a $\langle D_\mathrm{low-\ell}D_\mathrm{high-\ell}D_\mathrm{high-\ell}\rangle$ dust bispectrum contribution to $C_\ell^{\nu\kappa}$, that could bias our results.
{
We expect this contribution to be negligible as dust power is small in the high-$\ell$s used in the SMICA~\cite{cardosoComponentSeparationFlexible2008} foreground-cleaned CMB maps used in the PR4 lensing reconstruction.
Nevertheless, we test for such a bias here.
}
{If it were relevant then one would expect a larger effect on the Dustier sky region.}
Below $\ell \sim 100$, power in $C_{\ell}^{\nu \nu'}$ for the Dustier region is large than in the Cleaner region by a factor of 2, increasing to 5 around $\ell \sim 20$.
One might therefore estimate the bispectrum contribution to be~$2^{3/2}-5^{3/2}$ times larger in the Dustier region.
This would produce inconsistencies between it and the Cleaner region.
Additionally, as the dust power increases at lower $\ell$ we might expect to see inconsistencies in results as we vary the $\ell_{\mathrm{min}}$ used in the analysis.
This latter point is addressed in Fig.~\ref{fig:lmin_scaling}, where we can see that our \fnl{} posteriors are consistent as we increase $\ell_\mathrm{min}$, suggesting negligible dust bias.
For the former point we now consider the consistency of the two regions.

We can perform a null-test by considering the~$C_\ell^{\nu\kappa}$ signals on each area separately.
We consider the cross-power spectrum data vectors $ d^{(r)} \equiv \{ C_\ell^{\nu\kappa} \}_{\ell,\nu} $ constructed separately for each region $ r \in \{\mathrm{Cleaner}, \mathrm{Dustier}\} $.

We define the difference data vector
\begin{equation}
\Delta \equiv d^{(\mathrm{Cleaner})} - d^{(\mathrm{Dustier})}.
\end{equation}
Its covariance follows from standard linear propagation:
\begin{equation}
\mathrm{Cov}(\Delta)
= \mathbb{C}^{(\mathrm{Cleaner})}
+ \mathbb{C}^{(\mathrm{Dustier})}
- 2\,\mathbb{C}^{(\mathrm{Cleaner},\mathrm{Dustier})}.
\end{equation}
Here, $ \mathbb{C}^{(\mathrm{Cleaner})} $ and $ \mathbb{C}^{(\mathrm{Dustier})} $ correspond to the diagonal blocks of the full covariance matrix (i.e.~the two $3\times 3$ blocks in Fig.~\ref{fig:corr_mat}), while $ \mathbb{C}^{(\mathrm{Cleaner},\mathrm{Dustier})} $ is the off-diagonal block encoding cross-covariances between the two regions (as mentioned before, this is essentially zero).
We then construct $\chi^2_\mathrm{diff}$ as
\begin{equation}
\chi^2_{\mathrm{diff}} = \Delta^\intercal \, \mathrm{Cov}(\Delta)^{-1} \, \Delta,
\end{equation}
which, under the null hypothesis that the two data vectors are consistent realizations of the same underlying signal, is approximately distributed as a chi-squared variable with $ N $ degrees of freedom, where~$N$ is the length of our data vector.
For our data vectors $N=42$.
Evaluating this statistic, {and by comparing to a $\chi^2$ distribution with 42 degrees of freedom, $\chi_{42}^2$,} we obtain a $p$-value
\begin{equation}
p\big(\chi^2_{42} \ge \chi^2_{\mathrm{diff}}\big) = 0.37,
\end{equation}
indicating no statistically significant difference between the Cleaner and Dustier region data vectors.

Secondly, we can also consider the consistency through our constraints on \fnl{} in the two independent sky regions.
Fig.~\ref{fig:posterior_fnl} shows that the Cleaner and Dustier regions produce consistent constraints on \fnl{} to within $1\sigma$.
We also note that in \McCCIB{}, no evidence was found for any dependence of the CIB-$\kappa$ correlation on dust removal algorithm or sky region for a wide range of different measurements.

\subsection{Frequency dependence}\label{sec:frequency_dependence}

While the different frequency bands of the CIB are highly correlated, they are sourced by slightly different redshift kernels.
This means that the impact of~\fnl{} on the angular power spectrum of each will be different.

We therefore wish to explore the dependence of our \fnl{} constraints on the different frequencies used.
{The tests we discuss below were performed on both the real input data, and on mock spectra produced from theory.
The former allows us to test the stability of frequency choice, consistency or contamination within the data, or model misspecification.
The latter provides a cleaner exploration of the constraining power of the model across frequencies.
When we compared the results of these test we saw consistency results between both the data and mock versions.
Thus, here we only display the data versions of the tests, as they contain all of the effects.}

{To look at the contribution from each frequency channel separately, we first look at constraints when only a single frequency is used for $C_{\ell}^{\nu \kappa}$, and then when only two frequencies are used (See Appendix~\ref{app:isolated_freq_tests} for the full 2-dimensional posterior distributions).}
We keep all channels of $C_{\ell}^{\nu \nu'}$ as they provide negligible constraining power on \fnl, but are required for degeneracy breaking between nuisance parameters.

{For the single frequency case we find that the constraining power of each frequency alone is far reduced compared to the combined result.
Adding in a single additional frequency significantly reduces the width of the \fnl{} posterior distribution.
Most of this effect appears to be as a result of degeneracy breaking between nuisance parameters and \fnl{}(see Appendix~\ref{app:isolated_freq_tests}).}

In addition to frequency isolation, we also explore the effect inter-frequency correlations have on our constraints.
We, intriguingly, see some improvement from including the correlations in our covariance matrix.
These effects were tested using input theory vectors generated using the maximum a posterior parameters of the fiducial analysis.
Runs using a fully diagonal covariance matrix (main diagonal only of Fig.~\ref{fig:corr_mat}) see the \fnl{} posteriors inflated by $\sim 80\%$ when compared to the same theory vectors with the full covariance matrix.
{We show the full 2-dimensional posterior distributions for this case in Appendix~\ref{app:inter_freq_corr}}

The reduction in \fnl{} posterior width on inclusion of inter-frequency correlations could be from a number of effects.
The correlations could be providing information that effectively cleans additional dust from the maps.
If the correlation of the true CIB across frequency is known, excess foreground dust that is not correlated in the same way will have less impact on the likelihood, making us overall less sensitive to remaining dust.
Alternatively the inter-frequency correlations may be effectively removing the low-redshift information, where we are less sensitive to \fnl.
In the two-frequency case {(see Appendix~\ref{app:isolated_freq_tests})} we see that the case with no $545$ GHz channel has almost no change in posteriors from the full combination of frequencies.
This may suggest that some of the improvement gained on the $353$ alone constraints is from pinning down the lower redshift contribution with the $857$ GHz channel.

\section{Conclusions}\label{sec:conclusions}
We have presented constraints on the local-type primordial non-Gaussianity parameter, $f_\mathrm{NL}^\mathrm{local}$, derived from measurements of the scale-dependent bias in the cross-correlation between the CIB and the CMB lensing potential.
Using the new large-scale, galactic-dust–cleaned CIB maps of \McCCIB, which retain unbiased power down to lower multipoles than those used in previous analyses, we obtained significantly improved constraints compared to \McCfNL.
By combining multiple sky regions in a unified analysis for the first time, we find $f_\mathrm{NL}^\mathrm{local} = 43 \pm 23$, representing an overall factor of $\sim 2$ improvement in posterior width over~\McCfNL.
This cross-correlation alone constraint on \fnl{} is competitive with other recent CMB lensing cross-correlation studies (e.g.~\cite{FabbianConstraintsonprimordial2025, chiarenza_constraining_2025}) using different biased tracers,  which  see $\sigma(f_\mathrm{NL})\sim 22$.

The improvement with respect to \McCfNL{} is primarily driven by access to larger angular scales (lower $\ell_\mathrm{min}$) and by a more optimal joint treatment of sky regions with differing levels of galactic dust contamination.
We also found that the inclusion of correlation between CIB frequency improves the constraining power on \fnl.
This is not likely to be from degeneracy breaking alone, as we still see this effect when including the multiple frequencies of the CIB as if they were uncorrelated.
The specific cause of this cancellation will be investigated in future work.

Looking forward, further improvements in the constraints on \fnl{} may be achieved by the inclusion of better lensing measurements, including from ground-based CMB data.

\section*{Acknowledgements}
We thank Niall MacCrann, Abhi Maniyar, Yogesh Mehta, and Alexander van Engelen  for useful discussions. Computing was performed using resources provided through the STFC DiRAC Cosmos Consortium and hosted at the Cambridge Service for Data Driven Discovery (CSD3). JT was supported by the University of Cambridge Harding Distinguished Postgraduate Scholars Programme. We acknowledge support from the European Research Council (ERC) under the European Union’s Horizon 2020 research and innovation programme (Grant agreement No. 851274). The Flatiron Institute is a division of the Simons Foundation.

\bibliography{paper_references}

\appendix

\section{Simulations for covariance estimation}\label{app:sims}
For computation of the covariance matrix, and for the transfer function required to correct the misnormalisation of the masked lensing reconstruction (see Section~\ref{sec:MCNorm}), dust-contaminated CIB simulations that match the statistics of the measured maps are required.
These will need to include half-mission splits of the $R_N=16$ CIB maps for use in auto- and cross-frequency spectra, and full-mission simulations of the $R_N=1$ maps.
In addition, these simulations must be correlated with simulations of the CMB lensing potential, with the correct amount of foreground dust in the different $n_\mathrm{HI}$ threshold regions.

The base of these simulations are constructed from 480 NPIPE lensing simulations from Ref.~\cite{carronCMBLensingPlanck2022a}, each with a pair of input and reconstructed versions.
For a given input $\kappa$ simulation defined by a set of spherical harmonic coefficients $\{\kappa_{\ell m}\}$, one can construct a correlated CIB simulation as
\begin{equation}
    a^{\nu}_{\ell m} = f^\nu_\ell \kappa_{\ell m} + n_{\ell m}^\nu,\label{eq:corr_sims}
\end{equation}
where $\{a^\nu_{\ell m}\}$ are the set of spherical harmonic coefficients for the CIB field at frequency $\nu$.

The filter, $f^\nu_\ell$, converting the CMB lensing field to the correlated part of the CIB, and the contribution to the CIB uncorrelated with CMB lensing, $n_{\ell m}^\nu$, are defined such that
\begin{equation}
    \langle a^\nu_{\ell m} \kappa_{\ell m}\rangle = C_\ell^{\nu\kappa} \quad ; \quad \langle a^\nu_{\ell m} a^{\nu'}_{\ell m}\rangle = \hat{C}_\ell^{\nu\nu'} .\label{eq:sim_expectations}
\end{equation}
We achieve simulations that meet these definitions by taking $f_\ell^{\nu}$ as
\begin{equation}
    f^\nu_\ell = \frac{C_\ell^{\nu\kappa}}{C_\ell^{\kappa\kappa}}\label{eq:flnu},
\end{equation}
where $C_\ell^{\nu\kappa}$ is a fiducial theory prediction for the cross-correlation and  $C_\ell^{\kappa\kappa}$ is the theory curve used to generate the 480 lensing simulations.
The latter is taken from the mean of the angular power spectrum measured from the input simulations (pre-reconstruction) on the full sky.
We then draw an additional set of Gaussian simulations $n^{\nu}_{\ell m}$, such that
\begin{equation}
    \langle n_{\ell m}^\nu n_{\ell m}^{\nu'} \rangle = \hat{C}_\ell^{\nu\nu'} - \frac{C_\ell^{\nu\kappa}}{C_\ell^{\kappa\kappa}}\frac{C_\ell^{\nu'\kappa}}{C_\ell^{\kappa\kappa}}C_\ell^{\kappa\kappa}.\label{eq:nlm_expectations}
\end{equation}

We use measured spectra (indicated by the $\hat{}$ symbol) for the CIB auto-spectra so that the simulations match the observed amount of inter-frequency correlations.
This is taken as a smoothed version of $\hat{C}_{\ell}^{\nu \nu'}$ measured from the $R_N=16$ maps.
These maps are used as they have the highest amount of foreground dust removed, however they also remove some of the CIB signal on large scales.
To add this back we apply a low-$\ell$ transfer function to reapply some of this signal.
We estimate this transfer function from Fig. 4 of \McCCIB, with an additional $\propto1/\ell$ for $\ell\le10$ to ensure $C_{\ell}^{\nu\nu'}$ is large enough that $\frac{C_{\ell}^{\nu \kappa}}{\sqrt{ C_{\ell}^{\nu \nu'}C_{\ell}^{\kappa \kappa} }}>1$, as is required to draw the simulations.
The specifics of this transfer function are unimportant provided that when the dust is added the total power matches the measured spectra.
In Equations~\eqref{eq:sim_expectations}-\eqref{eq:nlm_expectations} $C_\ell^{\nu\kappa}$ is obtained from theory.
The~$C_\ell^{\nu\kappa}$ used for generating the simulations was taken from the best-fitting theory of when using a theoretical covariance, estimated from the fiducial parameters listed in Table~\ref{tab:priors}.
The relevant simulation sets are constructed by first constructing the background CIB simulations according to Equation~\eqref{eq:corr_sims}.
This is implemented through the use of \pypackage{healpy}'s \pypackage{synalm} function, which generates a set of $N$ correlated Gaussian random fields for $\frac{1}{2}N(N+1)$ input cross-spectra.

On top of the background CIB, foreground dust power is then accounted for by adding another set of Gaussian simulations.
The dust power is estimated from the difference between the $\hat{C}_{\ell}^{\nu \nu'}$ measured on the $R_{N}=1$ maps, and the simulated background CIB input $\hat{C}_{\ell}^{\nu \nu'}$ discussed above.
The estimated dust power for then Cleaner region is added first to the combined sky region defined by the $n_\mathrm{HI}<4.0\times 10^{20} \mathrm{cm}^{-2}$ boolean mask.
The difference between the estimated dust power on the Dustier and Cleaner regions is then added to the Dustier region alone.
Finally, two independent realizations of instrumental noise are generated, appropriate for half-mission splits.
These are then added to copies of the simulations described thus far, to produce two mock half-mission split simulations used in the CIB auto-spectra.
The half missions splits are coadded for a full mission simulation for all other spectra.

\begin{figure}
    \centering
    \includegraphics[width=\linewidth]{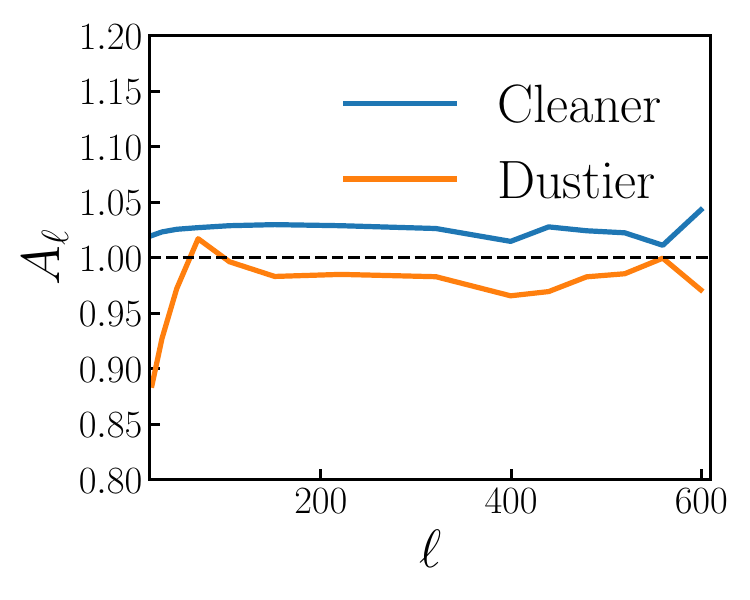}
    \caption{Transfer function to correct for the misnormalization of the reconstructed lensing maps as a result of masking.
    This is discussed in Section~\ref{sec:MCNorm} and Section~\ref{sec:cib_model}.
    The curves show the transfer function for the Cleaner (blue) and Dustier (orange) regions, as a function of $\ell$.}
    \label{fig:transfer}
\end{figure}

The full correlation matrix estimated from these simulations is shown in Fig.~\ref{fig:corr_mat}, and the MC normalization correction transfer function is shown in Fig.~\ref{fig:transfer}.

\section{2-dimensional posterior for the fiducial quadratic bias model}\label{app:corner_main}
\begin{figure*}
    \centering
    \includegraphics[width=0.9\linewidth]{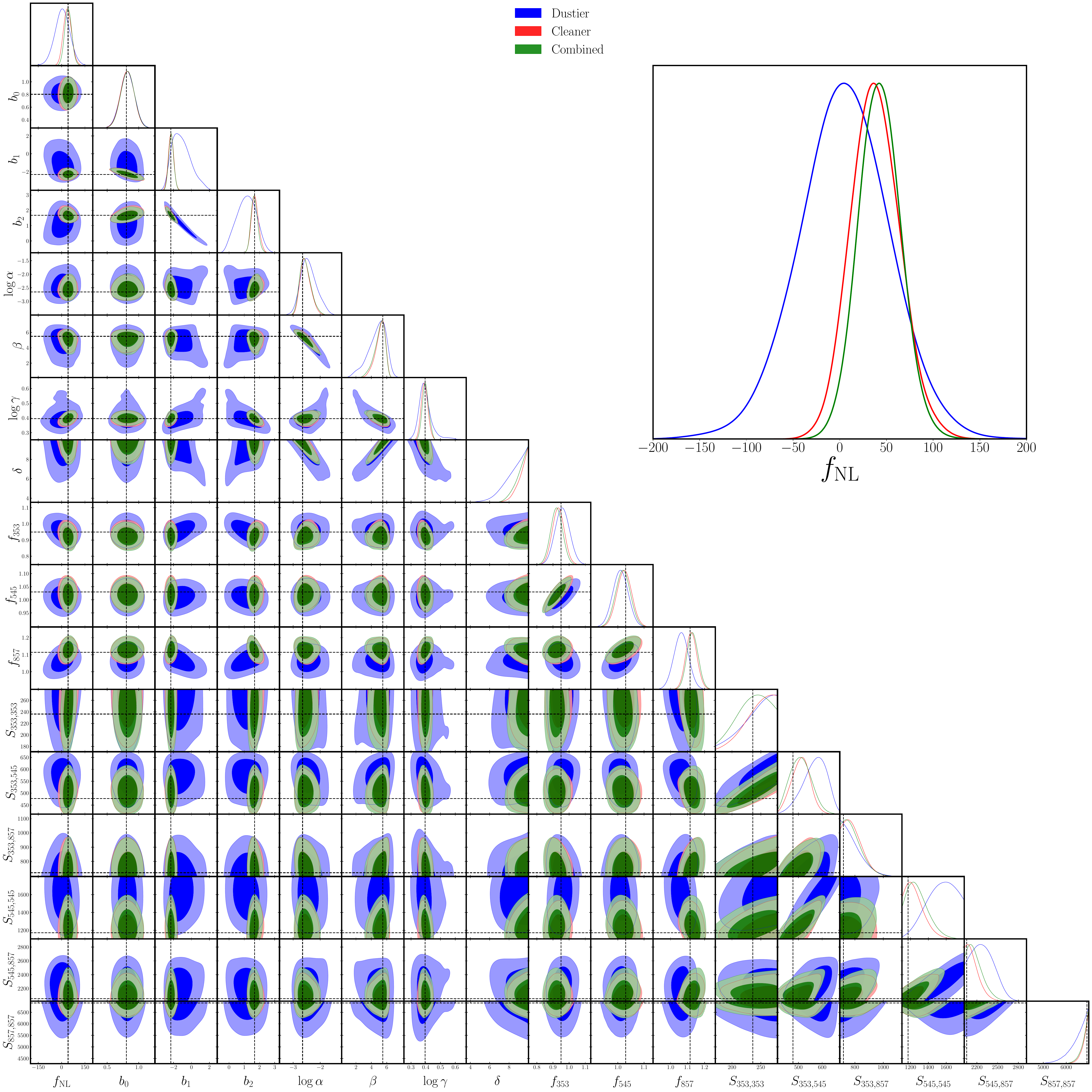}
    \caption{Corner plot of full parameter space for the two separate regions, the Cleaner region in red, and the Dustier in green, and their combined analysis in blue.
    {We find consistency between the two regions, and our tightest constraint is from their combination.}
    }
    \label{fig:corner_main}
\end{figure*}

Fig.~\ref{fig:corner_main} shows the fully marginalized 2-dimensional posteriors for the quadratic bias model, including the posteriors for the Cleaner, Dustier and combined region. 
We can clearly see that the Cleaner region has far higher constraining power on all parameters and that the posterior for the combined region is dominated by the Cleaner region.

\section{Isolated frequency tests}\label{app:isolated_freq_tests}
\begin{figure*}
    \centering
    \includegraphics[width=\linewidth]{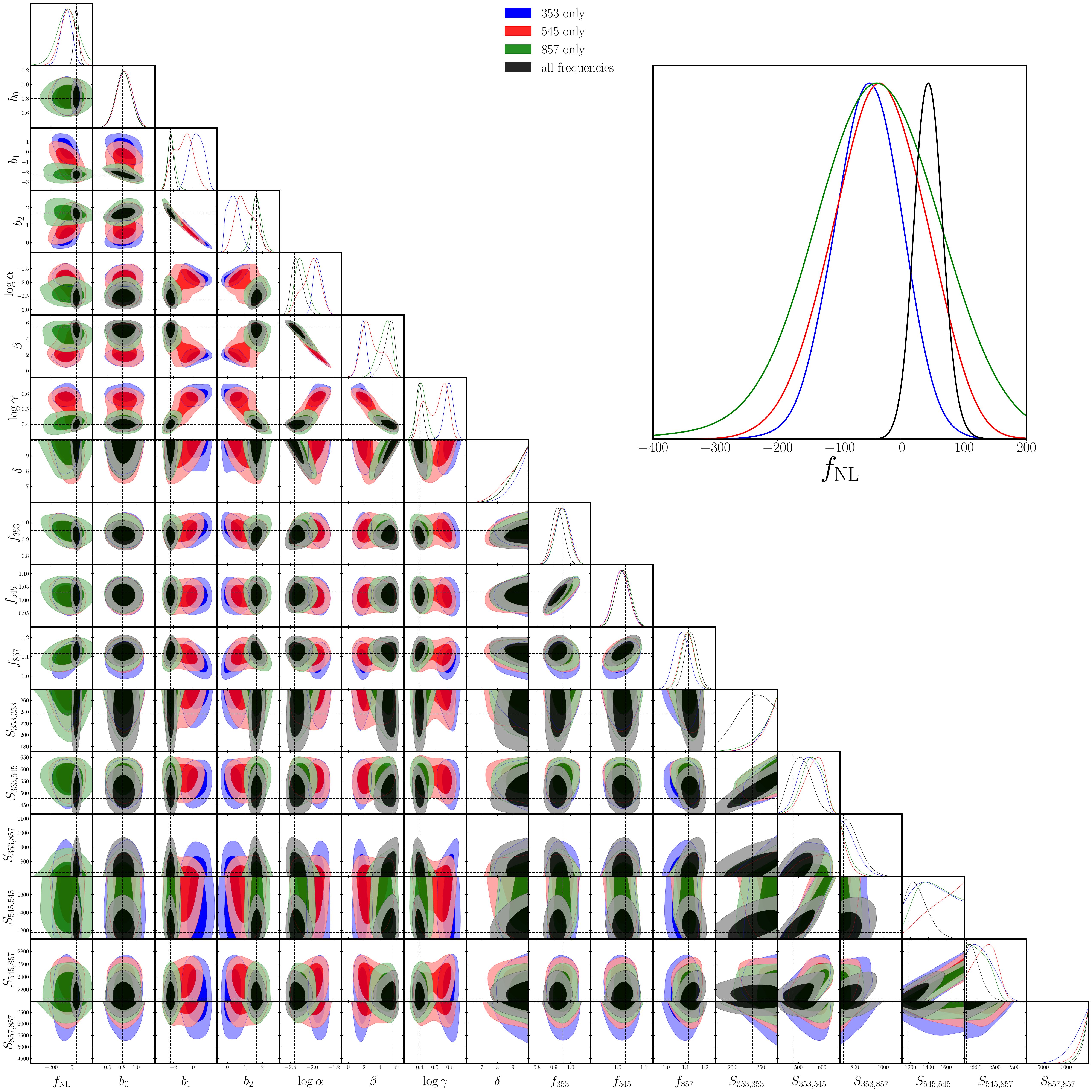}
    \caption{Corner plot of the full parameter space, when isolating a single frequency.
    The black lines indicate the maximum a posteriori point in parameter space.
    This plot demonstrates the degeneracy breaking upon inclusion of multi-frequency information.
    This is most easily seen in the $b_1$ vs \fnl{}, or $b_2$ vs \fnl{} panels.}
    \label{fig:single_freq_triangle}
\end{figure*}

\begin{figure*}
    \centering
    \includegraphics[width=\linewidth]{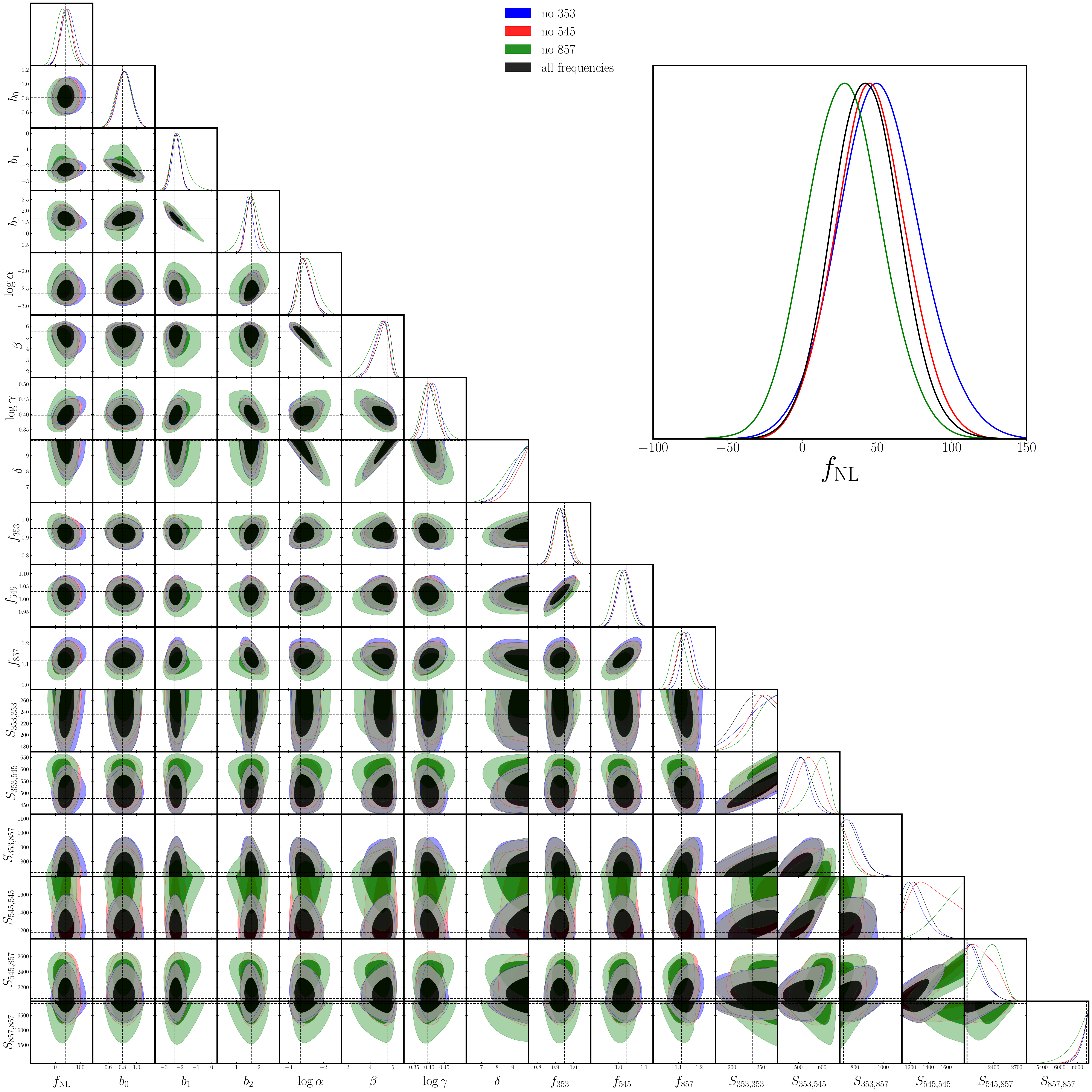}
    \caption{Corner plot of the full parameter space, when isolating two of the frequencies.
    The black lines indicate the maximum a posterior point in parameter space.
    {We find that all combinations of two frequencies produce consistent posterior distributions.}}
    \label{fig:double_freq_triangle}
\end{figure*}

Here we show the effect of running our analysis on single-frequency or double-frequency versions of our analysis compared to the full three-frequency version.
Fig.~\ref{fig:single_freq_triangle} shows the contours when we use only a single frequency, and Fig.~\ref{fig:double_freq_triangle} shows to the same for double frequencies.
We can see in Fig.~\ref{fig:single_freq_triangle} that the constraining power of each frequency alone is far reduced compared to the combined result.
{
Adding in a single additional frequency significantly reduces the width of the \fnl{} posterior distribution, as we can see in Fig.~\ref{fig:double_freq_triangle}.
Most of this effect appears to be as a result of degeneracy breaking between nuisance parameters and \fnl, and is most apparent in Fig.~\ref{fig:single_freq_triangle} when considering the change from single frequencies to the full analysis.
In Fig.~\ref{fig:single_freq_triangle} the degeneracy breaking is most apparent between the bias parameters ($b_1$ or $b_2$) and \fnl.
This degeneracy breaking also causes the central value of the \fnl{} posterior to move away from all of the single frequency posteriors.
This is again most aparent in the $b_1$ vs \fnl{} panel of Fig.~\ref{fig:single_freq_triangle}.
}

\section{Including inter-frequency correlations}\label{app:inter_freq_corr}
\begin{figure*}
    \centering
    \includegraphics[width=\linewidth]{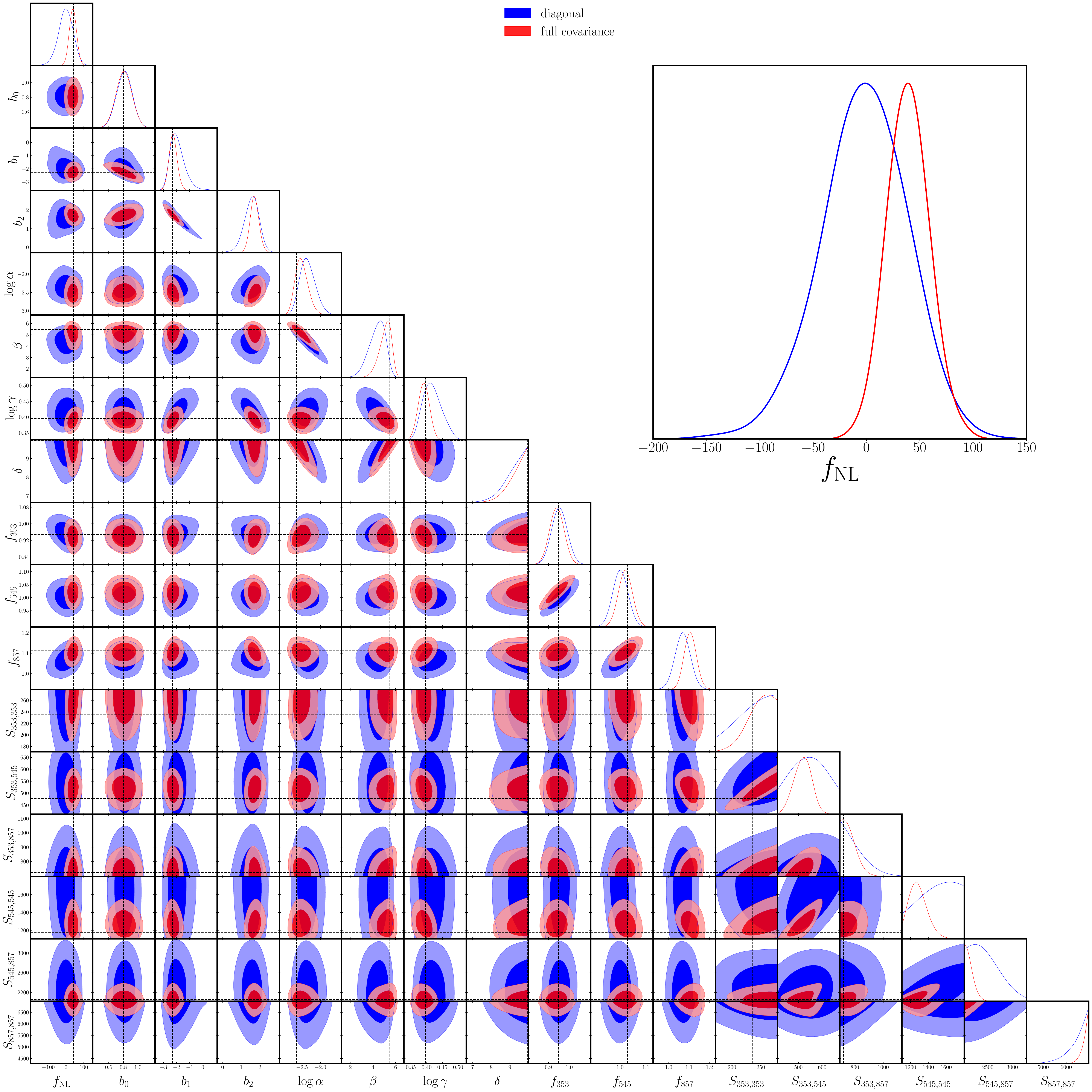}
    \caption{Corner plot of the full parameter space, when the covariance matrix included either contains correlation between frequency (off diagonals) or not.
    This is for an input theory data vector for the quadratic bias model in both cases, generated from the maximum a posteriori point in the parameter space for the real data.
    The black lines indicate the input theory parameters.
    We find that removing inter-frequency correlations significantly inflates the width of the \fnl{} posterior distribution.
    }
    \label{fig:comparison_triangle_full_full covmat_diagonal covmat}
\end{figure*}

Here we show the effect of running our analysis without including all known correlations between the data.
This is done by using only the main diagonal of the covariance matrix, so that there are no inter-frequency correlations included.
We ran these tests on a mock theory vector constructed from the MAP of the fiducial analysis, including updates to the priors on $\nu\bar{I_\nu}$.
Fig.~\ref{fig:comparison_triangle_full_full covmat_diagonal covmat} shows the contours when we use the diagonal covariance matrix compared to the full covariance matrix.
{We also ran this analysis including the off diagonals of $C_\ell^{\nu\nu'}$ only, finding no change from the full diagonal case.}

\end{document}